\documentclass[sigplan,table,x11names,dvipsnames,10pt]{acmart}

\usepackage[utf8]{inputenc}
\usepackage{xcolor}

\usepackage{ifthen}
\usepackage{pifont}
\usepackage{xspace}
\usepackage[font={small},skip=1pt,belowskip=3pt,labelfont=bf]{caption}

\usepackage{graphicx}
\usepackage{subcaption}

\usepackage{paralist}
\usepackage{tikz}
\usetikzlibrary{
    shapes,
    shapes.misc}

\newcommand{\sys}{\textsc{Kollaps}\xspace}
\newcommand{\tc}{\texttt tc}
\newcommand{\ks}{\textsc{Kubernetes}\xspace}

\newboolean{showcomments}
\setboolean{showcomments}{false}
\ifthenelse{\boolean{showcomments}}
{ \newcommand{\mynote}[3]{
    \fbox{\bfseries\sffamily\scriptsize#1}
    {\small$\blacktriangleright$\textsf{\emph{\color{#3}{#2}}}$\blacktriangleleft$}}}
{ \newcommand{\mynote}[3]{}}

\def\BibTeX{{\rm B\kern-.05em{\sc i\kern-.025em b}\kern-.08emT\kern-.1667em\lower.7ex\hbox{E}\kern-.125emX}}

\usepackage{listings}
\definecolor{backcolour}{rgb}{0.95,0.95,0.92}
\lstdefinestyle{xml}{
  numberbychapter=false,
  language=XML,
  numbers=left,                   
  stepnumber=1,                   
  numbersep=4pt,                  
  numberstyle=\footnotesize,      
  frame=no,           		  
  xleftmargin=10pt,
  basicstyle=\small,
  identifierstyle=\color{green},
  keywordstyle=\color{blue}\ttfamily,
  stringstyle=\color{red}\ttfamily,
  morekeywords={encoding,
    xs:schema,xs:element,xs:complexType,xs:sequence,xs:attribute, 
    name, image, replicas, origin, dest, orig,
    latency, drop, upload, up, download, down, network, jitter, loss},
  emph={experiment, services, bridges, service, node, bridge, links, link, dynamic, schedule, action, time},
  emphstyle=\color{olive},
}
\lstdefinestyle{cpp}{
  numberbychapter=false,
  language=C++,
  frame=single,           
  basicstyle=\small,
  keywordstyle=\color{blue}\ttfamily,
  stringstyle=\color{red}\ttfamily,
  emphstyle=\color{olive}
}
\lstdefinestyle{dockerfile}{
  numbers=left,                   
  stepnumber=1,                   
  numbersep=4pt,                  
  numberstyle=\footnotesize,      
  frame=no,           		      
  xleftmargin=10pt,
  basicstyle=\small,
  identifierstyle=\color{black},
  keywordstyle=\color{blue}\ttfamily,
  stringstyle=\color{red}\ttfamily,
  numberbychapter=false,
  morekeywords={FROM, RUN, CMD, LABEL, MAINTAINER, EXPOSE, ENV, ADD, COPY,
    ENTRYPOINT, VOLUME, USER, WORKDIR, ARG, ONBUILD, STOPSIGNAL, HEALTHCHECK,
    SHELL},
  morecomment=[l]{\#},
  morestring=[b]"
}
\lstdefinestyle{yaml} {
    columns=fixed,
    numbers=left,                   
    stepnumber=1,                   
    numbersep=6pt,                  
    numberstyle=\footnotesize\ttfamily,      
    xleftmargin=11pt,
    identifierstyle=\color{black}\ttfamily,
    stringstyle=\color{purple}\ttfamily,
    keywords=[1]{experiment, services, bridges, links, dynamic, schedule, action},
    keywordstyle=[1]\color{blue}\ttfamily,
    keywords=[2]{name, image, replicas, latency, up, down, jitter, dest, loss, time},
    keywordstyle=[2]\color{NavyBlue}\ttfamily,
    keywords=[3]{orig},
    keywordstyle=[3]\color{blue!50!purple}\ttfamily
}
\usepackage{marvosym}
\usepackage{enumitem}
\usepackage[authormarkup=none,final]{changes}
\definecolor{brickred}{rgb}{0.8, 0.25, 0.33}
\definechangesauthor[name=shady, color=brickred]{SI}
\definechangesauthor[name=valerio, color=brickred]{VS}
\definechangesauthor[name=mm, color=brickred]{MM}
\setcopyright{acmcopyright}
\acmConference[EuroSys '20]{Fifteenth European Conference on Computer Systems}{April 27--30, 2020}{Heraklion, Greece}
\acmISBN{978-1-4503-6882-7/20/04}
\setcopyright{none}
\acmISBN{}
\renewcommand\footnotetextcopyrightpermission[1]{} 
\usepackage{tcolorbox}
\begin{document}

\title[\sys: Decentralized and Dynamic Topology Emulation]{\sys:\\Decentralized and Dynamic Topology Emulation}

\author{Paulo Gouveia}
\affiliation{
  \institution{U. Lisboa \& INESC-ID}
  \city{Lisbon}
  \country{Portugal}
}

\author{João Neves}
\affiliation{
  \institution{U. Lisboa \& INESC-ID}
  \city{Lisbon}
  \country{Portugal}
}

\author{Carlos Segarra}
\affiliation{
  \institution{University of Neuchâtel}
  \city{Neuchâtel}
  \country{Switzerland}
}

\author{Luca Liechti}
\affiliation{
  \institution{University of Neuchâtel}
  \city{Neuchâtel}
  \country{Switzerland}
}

\author{Shady Issa}
\affiliation{
  \institution{U. Lisboa \& INESC-ID}
  \city{Lisbon}
  \country{Portugal}
}

\author{Valerio Schiavoni}
\authornote{Corresponding author}
\orcid{0000-0003-1493-6603}
\affiliation{
  \institution{University of Neuchâtel}
  \city{Neuchâtel}
  \country{Switzerland}
}
\email{valerio.schiavoni@unine.ch}

\author{Miguel Matos}
\authornote{Corresponding author}
\orcid{0000-0001-6916-2866}
\affiliation{
  \institution{U. Lisboa \& INESC-ID}
  \city{Lisbon}
  \country{Portugal}
}
\email{miguel.marques.matos@tecnico.ulisboa.pt}

\renewcommand{\shortauthors}{Gouveia et. al}

\begin{abstract}
The performance and behavior of large-scale distributed applications is highly influenced by network properties such as latency, bandwidth, packet loss, and jitter. 
For instance, an engineer might need to answer questions such as: What is the impact of an increase in network latency in application response time? 
How does moving a cluster between geographical regions affect application throughput? 
How network dynamics affects application stability? 
Answering these questions in a systematic and reproducible way is very hard, given the variability and lack of control over the underlying network. 
Unfortunately, state-of-the-art network emulation or testbeds scale poorly (\emph{i.e.}, MiniNet), focus exclusively on the control-plane (\emph{i.e}., CrystalNet) or ignore network dynamics (\emph{i.e.}, EmuLab).

\sys is a fully distributed network emulator that address these limitations. 
\sys hinges on two key observations. 
First, from an application's perspective, what matters are the emergent end-to-end properties (e.g., latency, bandwidth, packet loss, and jitter) rather than the internal state of the routers and switches leading to those properties. 
This premise allows us to build a simpler, dynamically adaptable, emulation model that circumvent maintaining the full network state. Second, this simplified model is maintainable in a fully decentralized way, allowing the emulation to scale with the number of machines for the application.

\sys is fully decentralized, agnostic of the application language and transport protocol, scales to thousands of processes and is accurate when compared against a bare-metal deployment or state-of-the-art approaches that emulate the full state of the network.
We showcase how \sys can accurately reproduce results from the literature and predict the behaviour of a complex unmodified distributed key-value store (\emph{i.e.}, Cassandra) under different deployments.

\end{abstract}

\ccsdesc[500]{Networks~Network experimentation}

\keywords{distributed systems, emulation, dynamic network topology, containers, experimental reproducibility}

\maketitle
\thispagestyle{fancy}
\section{Introduction}
\label{sec:introduction}
Evaluating large-scale distributed systems is hard, slow, and expensive.
This difficulty stems from the large number of moving parts one has to be concerned about: system dependencies and libraries, heterogeneity of the target environment, network variability and dynamics, among others.

Such uncontrollable and poorly specified environment leads to a slow experimental cycle, results that are hard to reproduce, and potential downstream costs worth millions of dollars~\cite{bailis2014network}.
It is therefore of the utmost importance to have tools that allow to precisely describe the environment and control key parts of the system such as the network.

On the one hand, the advent of container technology (\emph{e.g.}, Docker~\cite{Merkel2014}, Linux LXC~\cite{lxc}) and container orchestration (\emph{e.g.}, Docker Swarm~\cite{dockerswarm}, Kubernetes~\cite{k8s}) greatly simplifies the description and deployment of complex systems and partially addresses the problem.
On the other hand, there is an acute need for tools that allow to precisely control the network in complex, large-scale experiments. 

As a matter of fact, the inherent variability of WAN conditions (\emph{i.e.}, failures, contention and reconfigurations), makes it hard to assess the impact of changes in the application logic.
Is the observed performance improvement really due to improvements in the application, or due to a \emph{lucky} run when the network was lightly loaded?
How is performance affected by common network dynamics, such as background traffic, or link flapping?
The very same questions and issues also arise in the reproducibility crisis currently plaguing \deleted[id=SI]{down} the system's community~\cite{Boisvert:2016:IR:3001840.2994031,peng2011reproducible}.
Different results for the same system emerge not only because systems are evaluated in different uncontrollable conditions, but also because research testbeds such as Emulab~\cite{hibler2008large}, CloudLab~\cite{ricci2014introducing}, or PlanetLab~\cite{chun2003planetlab} used to conduct experiments tend to get overloaded right before system conference deadlines~\cite{Kim:2011:UCP:2068816.2068864}.
We therefore need tools to systematically assess and reproduce the evaluation of large-scale applications.
\replaced[id=SI]{Notably}{Noteworthy}, similar tools could be used to test and (otherwise) prevent costly incidents due to mis-configurations, as in recent events~\cite{gcspostmortem,awsdynamopostmortem}.

One approach to systematically evaluate a large-scale distributed system is to resort to simulation, which relies on models that capture the key properties of the target system and environment~\cite{banks2010discrete}. 
Simulation provides full control of the system and environment --- achieving full reproducibility --- and allows to study the model of the system in a variety of scenarios.
However, simulations suffer from several well-known problems.
In fact, there is a large gap between the simulated model and the real-world deployment, usually leading to several unforeseen behaviors not captured by the model~\cite{Chandra:2007:PML:1281100.1281103,floyd2003internet,floyd2001difficulties,Paxson97whywe}.
And even if the simulated model yields correct results, the real implementation is not guaranteed to faithfully follow the simulated model.
Moreover, despite some efforts to model complex systems either through formal method analysis~\cite{Newcombe:2015:AWS:2749359.2699417} or simulation, this is, to the best of our knowledge, seldom the case for large-scale systems.

The alternative is to resort to network emulation.
In a network emulation, the real system is run against a model of the network that replicates real-world behavior by modeling a network topology together with its network elements, including switches, routers and their internal behavior.
Emulation thus allows to reach conclusions about the behavior of the real system in a concrete scenario rather than its model.
Unfortunately, state-of-the art network emulators suffer from several limitations.
MiniNet~\cite{handigol2012mininet} is limited to a single physical machine and therefore cannot be used to emulate a large-scale resource-intensive system.
MaxiNet~\cite{Wette2014} and the multi-host version of MiniNet support distributed clusters but scale poorly~\cite{liu2017crystalnet}. 
ModelNet~\cite{Vishwanath2009} and alike rely on a dedicated cluster of nodes to maintain the emulation model to which the application nodes must connect.
However, accuracy is \deleted[id=SI]{highly} dependent on application traffic patterns and \replaced[id=SI]{can degrade}{quickly degrades} with a modest increase in the number of application nodes\added[id=MM]{~\cite{modelnet}}.
CrystalNet~\cite{liu2017crystalnet} accurately emulates the \emph{control-plane} of large-scale networks (\emph{e.g.} routing tables, software switch versions or device firmwares) but cannot be used to emulate the data-plane (\emph{e.g.} latency, bandwidth), and hence evaluate the behavior of large-scale distributed applications.
While emulation testbeds (\emph{e.g.}, Emulab~\cite{hibler2008large}) provide a semi-controlled environment and network, they cannot model network dynamics and thus one cannot assess their impact on application behavior.
In summary, existing tools cannot systematically assess and reproduce the evaluation of large-scale applications subject to network dynamics.
\newcommand{\YES}{\textcolor{NavyBlue}{\ding{51}}}
\newcommand{\NO}{\color{BrickRed}{\ding{55}}} 
\newcommand{\DK}{{\color{BlueViolet} --\xspace}}
\newcommand{\DYN}{{\color{Red} \text{\Lightning}\xspace}}


\begin{table*}[t!]
    \caption{Classification of network emulation tools. NetEm~\cite{netem} uses a different queueing discipline to implement bandwidth shaping. Dockemu uses ns-3~\cite{riley2010ns} for link-level features. \DYN: ability to dynamically change this property. P=process,  V=virtual machine, C=container.}
\setlength{\tabcolsep}{3pt}
  \scriptsize
  \center
  \rowcolors{1}{gray!10}{gray!0}
  \begin{tabular}{r|c|c|c|c|c|c|c|c|c|c|c|c|l}
  \rowcolor{gray!25}
  \textbf{} & \textbf{} & \textbf{} & \textbf{} & \textbf{} & \textbf{Concurrent} & \textbf{Path} & \multicolumn{4}{c|}{\textbf{Link-Level emulation capabilities}} & \textbf{Any} & \textbf{Topology} &\textbf{}\\
  \rowcolor{gray!25}
  \textbf{Name} & \textbf{Year} &  \textbf{Mode} & \textbf{HW ind.} & \textbf{Orchestration} & \textbf{deployments} & \textbf{congestion} & \textbf{Bandwidth} & \textbf{Delay} & \textbf{Packet loss} & \textbf{Jitter} & {\textbf{Language}} & \textbf{dynamics} &{\textbf{Unit}}\\
  \hline
    DelayLine~\cite{ingham1994delayline}             & 1994 & User   & \YES  & Centralized     & \NO  & \NO  & \NO  & \YES & \YES & \NO  & \YES & \NO & P \\
  ModelNet~\cite{modelnet}                         & 2002 & Kernel & \YES  & Centralized     & \NO  & \YES & \YES\DYN & \YES\DYN & \YES\DYN & \NO  & \YES & \YES & P \\
  Nist NET~\cite{carson2003nist}                   & 2003 & Kernel & \YES  & Centralized     & \NO  & \NO  & \YES\DYN & \YES\DYN & \YES\DYN & \YES\DYN & \YES & \NO & P \\
    NetEm~\cite{netem}                               & 2005 & Kernel & \YES  & \multicolumn{3}{c|}{\scriptsize{\emph{(N/A: single link emulation only)}}} & \NO  & \YES & \YES & \YES & \YES & \NO & P\\
    Trickle~\cite{eriksen2005trickle}                & 2005 & User   & \YES  & \multicolumn{3}{c|}{\scriptsize{\emph{(N/A: single link emulation only)}}} & \YES & \YES & \NO  & \NO & \YES & \NO & P\\
  EmuSocket~\cite{avvenuti2006application}         & 2006 & User   & \YES  & \multicolumn{3}{c|}{\scriptsize{\emph{(N/A: single link emulation only)}}} & \YES\DYN & \YES\DYN & \NO  & \NO & \YES & \NO & P\\
    ACIM/FlexLab~\cite{sanaga2007flexlab}            & 2007 & Kernel & \YES  & Centralized     & \NO  & \YES & \YES\DYN & \YES\DYN & \YES\DYN & \YES\DYN & \YES & \YES & V \\
  NCTUns~\cite{wang2007design}                     & 2007 & Kernel & \YES  & Centralized     & \NO  & \YES & \YES & \YES & \YES & \YES & \YES & \NO & P \\
    Emulab~\cite{emulabOSR,hibler2008large}          & 2008 & Kernel & \NO   & Centralized     & \NO  & \YES & \YES\DYN & \YES\DYN & \YES\DYN & \NO & \YES & \YES & V \\
  IMUNES~\cite{puljiz2008performance}              & 2008 & Kernel & \NO   & Centralized     & \NO  & \NO  & \YES & \YES & \YES & \NO  & \YES & \NO & P \\
  MyP2P-World~\cite{roverso2008myp2pworld}         & 2008 & User   & \YES  & Centralized     & \NO  & \NO  & \YES & \YES & \YES & \NO  & \NO  & \NO & P \\
  P2PLab~\cite{nussbaum2008lightweight}            & 2008 & Kernel & \YES  & Centralized     & \NO  & \NO  & \YES & \YES & \YES & \NO  & \NO  & \NO & P \\
  Netkit~\cite{pizzonia2008netkit}                 & 2008 & Kernel & \YES  & Centralized     & \NO  & \YES & \YES\DYN & \YES\DYN & \YES\DYN & \NO  & \YES & \NO & V\\
    DFS~\cite{tang2009dsf}             		         & 2009 & User   & \YES  & Centralized     & \YES & \NO  & \YES\DYN & \YES\DYN & \YES & \YES & \NO  & \YES & P \\
    Dummynet~\cite{Carbone2010}                      & 2010 & Kernel & \YES  & Centralized     & \NO  & \NO  & \YES\DYN  & \YES\DYN & \YES\DYN & \NO & \YES & \NO & P \\
  Mininet~\cite{lantz2010network}                  & 2010 & Kernel & \YES  & Centralized     & \NO  & \YES & \YES\DYN & \YES\DYN & \YES\DYN & \YES\DYN  & \YES & \YES & P \\
  SliceTime~\cite{Weingartner:2011qy}              & 2011 & Kernel & \NO   & Centralized     & \NO  & \YES & \YES & \YES & \NO  & \NO  & \YES & \YES & V \\
  Mininet-HiFi~\cite{handigol2012mininet}          & 2012 & Kernel & \YES  & Centralized     & \NO  & \NO  & \YES\DYN & \YES\DYN & \YES\DYN & \YES\DYN & \YES & \YES & C \\
  \textsc{SplayNet}~\cite{schiavoni2013splaynet}	 & 2013 & User   & \YES  & Decentralized   & \YES & \YES & \YES & \YES & \YES & \NO  & \NO  & \YES & P \\
    MaxiNet~\cite{Wette2014}			         & 2014 & Kernel & \YES  & Centralized     & \NO  & \YES & \YES\DYN & \YES\DYN & \YES\DYN & \YES\DYN & \YES & \YES & P\\
    Dockemu~\cite{To2015}                            & 2015 & User   & \YES  & Centralized     & \NO  & \NO  & \YES & \YES & \YES & \YES & \YES & \NO & C \\
    EvalBox~\cite{evalBox2015}                       & 2015 & Kernel & \YES  & Centralized     & \NO  & \NO  & \YES\DYN & \YES\DYN & \YES\DYN & \YES\DYN & \YES  & \YES & P \\
    ContainerNet~\cite{peuster2018containernet}          & 2016 & Kernel & \YES  & Centralized     & \NO  & \YES & \YES\DYN & \YES\DYN & \YES\DYN & \YES\DYN  & \YES & \YES & C,V \\
    Kathar\'{a}~\cite{bonofiglio2018kathara}          & 2018 & Kernel & \YES  & Centralized     & \NO  & \YES & \YES\DYN & \YES\DYN & \YES\DYN & \NO  & \YES & \NO & C\\
    \hline
      \textbf{\sys}							         &   \textbf{2020}   & \textbf{Kernel} & \YES  & \textbf{Decentralized}   & \YES & \YES & \YES\DYN & \YES\DYN & \YES\DYN & \YES\DYN & \YES & \YES & \textbf{C,V} \\
  \end{tabular}
  \label{table:rw}
  \vspace{-10pt}
\end{table*}

In this paper we introduce \sys, a  \replaced[id=SI]{decentralized network emulator for large-scale applications}{network emulator that enables such experiments}.
\sys overcomes the state-of-the-art limitations through three key insights.
First, from the perspective of a distributed application, the observable end-to-end properties, such as latency, jitter, bandwidth and packet loss, are more relevant to its behavior than the underlying state of each networking element leading to these properties.
This enables us to build a simplified model that does not require emulating the full-state of the internal network elements (\emph{e.g.}, routers, switches) but provides equivalent behavior.
Second, it is possible to accurately maintain this emulation model in a fully-distributed fashion thus allowing the emulation to scale with the application nodes without sacrificing accuracy.
Finally, the simplified model lends itself to quick changes enabling us to emulate dynamic events such as link removals and additions or background traffic in a fraction of a second.
\begin{figure}[!t]
\includegraphics[scale=0.63]{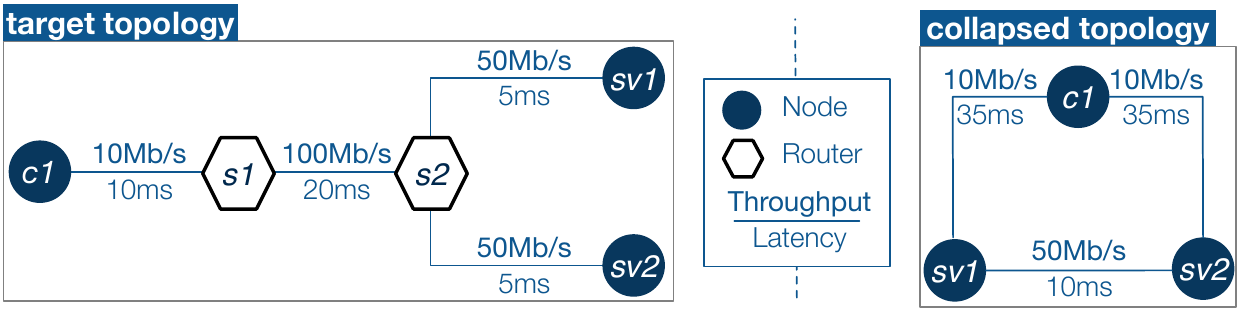}
\centering
\caption{Left: 3 application containers (\emph{c1}, \emph{sv1}, \emph{sv2}) and 2 network elements (\emph{s1}, \emph{s2}). Right: collapsed topology \added[id=SI]{with links describing maximum bandwidth and minimum latency between each two nodes}.}\label{fig:topology}
\end{figure}
To understand \sys's main principle, consider the topology in Figure~\ref{fig:topology} (left), with two network elements (\emph{s1} and \emph{s2}) and three containers (\emph{c1}, \emph{sv1}, and \emph{sv2}).
Rather than emulating the full network and the state of the switches (\emph{s1} and \emph{s2}), we rely on a \emph{network collapsing} technique to collapse the topology to virtual end-to-end links that retain the properties of the original topology, as depicted in Figure~\ref{fig:topology} (right). 
\added[id=SI]{Note that the bandwidth and latency values in the collapsed network depict the maximum available bandwidth and the minimum achievable latency between each two nodes, as if it is the only active flow. } The \added[id=SI]{actual} link properties are then maintained through a \emph{distributed network emulation} algorithm that models latency, bandwidth, jitter and packet loss.
The distributed nature of \sys{} and its simplified emulation model allows it to scale to thousands of processes with accuracy on par with centralized state-of-the-art emulators.

\textbf{Contributions.}
Our main contributions are:
\vspace{-5pt}
\begin{enumerate}[leftmargin=*]
\setlength\itemsep{0em}
\item \sys, the first network topology emulator that allows to evaluate large-scale applications in dynamic networks;

\item We integrate \sys with Docker Swarm~\cite{dockerswarm} and Kubernetes~\cite{k8s}, to deploy and evaluate unmodified containerized (distributed) systems;

\item A comparison of \sys's emulation accuracy versus bare-metal deployments and state-of-the-art approaches. Our evaluation scenarios includes static and dynamic topologies, various workload patterns (\emph{i.e.}, short and long-lived data flows) and different TCP congestion models, \added[id=VS]{including TCP Reno and Cubic}; 

\item A showcase of the new types of experiments that \sys enables. Namely, we reproduce results from published papers~\cite{Sousa2016,Bessani2014a}, and assess how Apache Cassandra~\cite{Lakshman:2010:CDS:1773912.1773922,cassandra} is affected by different network characteristics \emph{as-if} it were deployed on AWS. 
\end{enumerate}
\textbf{Organization.} This paper is organized as follows.
We survey related work in \S\ref{sec:relatedWork}.
The design and system architecture of \sys are described in \S\ref{sec:system}, with implementation details presented in \S\ref{sec:implementation}.
In \S\ref{sec:evaluation} we present our in-depth evaluation.
We discuss the current limitations and future work in \S\ref{sec:limitations}, before concluding in \S\ref{sec:conclusion}.

\section{Related Work}
\label{sec:relatedWork}

We categorize network emulators along several dimensions (Table~\ref{table:rw}): 
where the link shaping is executed (user/kernel mode),
independence from the underlying hardware,
type of orchestration across the cluster (centralized or decentralized),
support for concurrent experiments and users,
support for path congestion (\emph{i.e.}, multiple independent flows sharing the same emulated link), link-level emulation features (bandwidth, delay, packet loss, jitter),
ability to dynamically adjust such features on the fly as well as to change the topology itself
 (\emph{i.e.}, add/remove links, switches and nodes),
implementation-language restrictions for the programs under emulation,
and the supported deployment unit (virtual machines, containers or native processes).
Due to lack of space, we only detail how \sys compares with some representative systems, and we consider simulation-based tools (\emph{e.g.}, ns-3~\cite{riley2010ns}, PeerSim~\cite{montresor2009peersim}) out of the scope.


Few recent works cover orthogonal aspects of network emulation and illustrate the relevance of controlled experiments.
CrystalNet~\cite{liu2017crystalnet} focuses on large-scale emulation of the control-plane, enabling network engineers to evaluate changes to the control-plane before deploying in production.
\sys only deals with data-plane, and it is hence complementary to CrystalNet. \added[id=SI]{To enhance the efficiency of large-scale control-plane analysis, Bonsai~\cite{Beckett2018Bonsai} leveraged the idea of network compression while persevering the network properties. \sys' network collapsing achieves a similar goal, however, it targets a different set of network properties, again in the data-plane.} 
Pantheon~\cite{yan2018pantheon} is used to evaluate Internet congestion-protocols.
It gathers ground-truth data and compares it with results obtained from several emulators for a variety of congestion control algorithms.
The work done in Pantheon provides evidence that it is possible to approximate the behavior of a wide range of congestion algorithms by relying only on a small number of end-to-end properties.
In this paper, we rely on the same insight to provide a network emulator able to emulate large-scale topologies with accuracy.

\textbf{User-space approaches.} Trickle~\cite{eriksen2005trickle} uses dynamic linking and preloading functionality of Unix-based systems to insert its code between unmodified binaries and the system calls to the sockets API.
It performs bandwidth shaping and delay before delegating to the actual underlying socket calls, based on a simple configuration process.
Although multiple instances of Trickle can cooperate, setting up a multi-host system to emulate large networks involves tedious and error-prone manual configuration since there is no central deployment control system. 
Further, Trickle does not support statically linked binaries.
In contrast, \sys is independent of the application as it works with unmodified binaries, either dynamically or statically linked. 
EmuSocket~\cite{avvenuti2006application} and DelayLine~\cite{ingham1994delayline} are userspace tools, similar in design and features to Trickle.
DelayLine supports the deployment of complex topologies, but it lacks several important network emulation features (such as bandwidth or jitter).

MyP2P-World~\cite{roverso2008myp2pworld} is a Java-based application-level emulator aimed at peer-to-peer protocols. 
Applications must be implemented in Java and rely on Apache Mina~\cite{apache-mina} to intercept and emulate large-scale network conditions.
While \sys can be used for Java applications, it can be used with any other language as well.

\textsc{SplayNet}~\cite{schiavoni2013splaynet} extends \textsc{Splay}~\cite{Leonini2009} to allow emulation of arbitrary network topologies, deployed across several physical hosts in a fully decentralized manner. 
\textsc{SplayNet}, is fully distributed as it does not rely on  dedicated processes for network emulation. 
To emulate the network topology, \textsc{SplayNet} relies on graph analysis and distributed emulation algorithms, effectively collapsing the inner topology and delivering packets directly from one emulated host to the destination host.
However, it requires developers to implement their programs in a Domain Specific Language using the Splay framework and the Lua programming language, precluding its usage to evaluate real-world systems.
Moreover, it does not support dynamics nor does it emulate packet loss upon congestion.
\sys adopts a similar fully decentralized approach while completely overcoming its limitations. 
In fact, \sys can be used with unmodified, off-the-shelf applications and assess their performances under different network conditions also including dynamic topologies.

\textbf{Kernel-space approaches.} Next, we survey network emulators that require explicit or specialized support from the underlying OS and kernel.
Dummynet~\cite{Rizzo1997} operates directly on a specific network interface.
It is used as a low-level tool to build full-fledged emulators, such as Modelnet~\cite{modelnet}.
Modelnet allows the deployment of unmodified applications.
Applications are deployed on \emph{edge} nodes and all network traffic is routed through a set of \emph{core routers} - dedicated machines that collectively emulate the properties of the desired target network before relaying the packets back to the destination's edge nodes.
\sys relies on Linux's Traffic Control (\texttt{tc})~\cite{linuxtc} to offer \replaced[id=VS]{the same}{similar} low-level traffic shaping features, but (1) without requiring dedicated hosts and (2) at the same time providing a complete testbed integrating with large-scale container orchestration tools. 

The Emulab~\cite{emulabOSR} testbed supports the deployment of user-provided operating systems. 
As ModelNet and \sys, it leverage Linux's \texttt{tc} to shape the traffic directly at the edge nodes.
Emulab supports large topologies over shared clusters while maintaining the user requested resource allocation, and the ability to perform this scheduling optimally.
Its graph coarsening technique is similar in principle to \sys approach for collapsing the topology.
%

\textbf{Container-based approaches.} Finally, we look at emulation tools used with containers.
Mininet~\cite{lantz2010network} emulates network topologies on a single host.
It relies on Linux's lightweight virtualization mechanisms (\emph{i.e.}, \texttt{cgroups}) to emulate separated network hosts.
Similarly to Docker, it creates virtual Ethernet pairs running in separated namespaces and it assigns processes to those.
Mininet can emulate hundreds of networked hosts (instances) on a single physical host, with dedicated instances for switches and routers running on their own processes.
Conversely, \sys does not require these additional network instances, relying instead on maintaining and updating the state of the emulation at each container.
Mininet is limited to a single-host deployment hence preventing its use for large-scale resource-intensive applications that cannot fit a single machine.
Besides, even with a simple topology, Mininet's accuracy quickly degrades under certain workloads such as short-flows. 
Maxinet~\cite{Wette2014} extends Mininet to allow for cluster deployments of worker hosts and with native support for Docker containers.
It does so by tunneling links that cross different workers. 
However, it requires all emulated hosts that connect to the same switch to be deployed on the same worker as the switch.
In contrast, \sys does not impose co-located deployments of workers and switches.
ContainerNet~\cite{peuster2016medicine,peuster2018containernet} extends Mininet to add native support for Docker containers and dynamic topologies.
Still, it is limited to single machine deployments.
A similar limitation is present in Dockemu~\cite{To2015}, a network emulation tool based on Docker containers.

To the best of our knowledge, \sys is the only system that can be used to evaluate unmodified large-scale applications over arbitrary topologies, supporting a richer set of emulation features, and providing good accuracy when compared to  bare metal and state-of-the-art systems. \added[id=SI]{Finally, it is worth noting that the decentralized design of \sys' metadata exchange is similar in spirit to the 
hose model of ElasticSwitch~\cite{Popa2013ElasticSwitch}, which was used to provide bandwidth guarantees for virtual machines in cloud environments.}
\begin{figure}[t]
\centering
\includegraphics[scale=0.9]{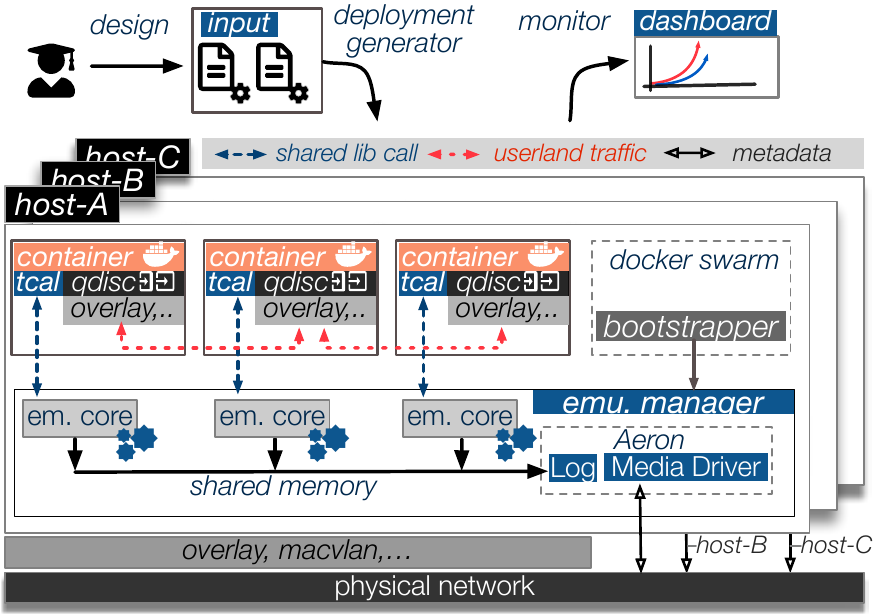}
\caption{\sys system design. Note that the \emph{bootstrapper} component is not required in a Kubernetes deployment (\S\ref{sec:implementation}).\label{img:arch}}
\end{figure}

\section{The \sys System}
\label{sec:system}

In this section we describe the system design and architecture of \sys. 
Figure~\ref{img:arch} depicts the main components and a deployment over several hosts. 
\sys consists of several components. 
The Emulation Manager has a single instance per physical machine and is responsible for maintaining and enforcing the emulation model. 
The TC Abstraction Layer (TCAL) is deployed once per application container, and it retrieves and sets the link properties. 
The Bootstrapper is started once per physical machine: it initiates \sys in Docker Swarm deployments (but as detailed in \S\ref{sec:implementation} is not needed under Kubernetes).
The Dashboard exposes a web-based interface to monitor and control the experiments. 
Finally, the Deployment Generator converts an experiment description (\emph{e.g.}, Listing~\ref{kol:yaml:static} and Listing~\ref{kol:yaml:dyn}) into a deployment plan. 
Next, we detail these components and their interactions. 

The \textbf{Deployment Generator} translates a topology description into an actual container deployment plan. 
\sys supports an XML Modelnet-like syntax~\cite{modelnet} to facilitate porting of existing topology descriptions, as well as a lean YAML-based syntax that we show here. 
Listing~\ref{kol:yaml:static} describes the network topology from Figure~\ref{fig:topology} (left). 
The topology description language supports \texttt{services}, \texttt{bridges}, \texttt{links}, and \texttt{dynamic} elements. 
The \texttt{services} correspond to sets of containers sharing the same image. 
The image names must be valid and available from private or public Docker registries (Listing~\ref{kol:yaml:static}, lines 4 and 6). 
Each service supports several parameters (\emph{e.g.}, total replicas or additional parameters to pass to the running containers once deployed). 
The \texttt{bridges} map to networking devices, \emph{e.g.}, routers and switches that have unique names (lines 9--10) and are arbitrarily connected, to realize complex topologies, via \texttt{links}.
Links can be uni-- or bi---directional, with mandatory attributes to specify the source, destination, properties (\emph{i.e.}, latency, bandwidth, packet loss, jitter), and the name of the container network to attach to (lines 12--17).
In case of jitter, the link latency follows by default a normal distribution but others are supported, with mean and standard deviation to match the specified latency and jitter attributes.
Internally, all links are unidirectional: declaring a bi-directional link results in the creation of two identical links in opposite directions, with same attributes except for the bandwidth capacity where upload and download attributes might differ.
The \texttt{dynamic} part (Listing~\ref{kol:yaml:dyn}) injects changes into the topology dynamically while the experiment progresses.
\sys supports a rich set of dynamic events, \emph{e.g.}, modification of any of the properties of the links, addition and removal of links, bridges and services.
This captures a wide range of dynamics, not only in the application itself, whose nodes (containers) may come and go during the experiment, but also in the network topology. 
For example, the rapid removal and insertion back into the topology of a link emulates a flapping link~\cite{Potharaju:2013:NCE:2523616.2523638}.
Each event maps to an \texttt{action} element, either for changes to link properties (lines 21--23) or for addition and removal of services, links and bridges (lines 24--36).
\added[id=MM]{A dedicated DSL to easily program more complex dynamic patterns on top of \sys is detailed in~\cite{thunderstorm}.}


\begin{figure}[t!]
\noindent\begin{minipage}[t]{.45\columnwidth}
\begin{lstlisting}[style=yaml, caption=Static topology.,  captionpos=t,  label=kol:yaml:static]{Name}
experiment: 
 services: 
  name: c1
   image: "iperf"
  name: sv
   image: "nginx"
    replicas: 2
 bridges: 
  name: s1
  name: s2
 links: 
  orig: c1
   dest: s1
   latency: 10
   up: 10Mbps
   down: 10Mbps
   jitter: 0.25
#others not shown     
\end{lstlisting}
\end{minipage}\hfill
\hspace{-15pt}
\begin{minipage}[t]{.43\columnwidth}
\begin{lstlisting}[style=yaml, caption=Dynamic events., captionpos=t, label=kol:yaml:dyn, firstnumber=19]{Name}
dynamic: 
 orig: c1
  dest: s1
  jitter: 0.5
  time: 120    
 action: leave
  name: s1
  time: 200 
 action: join
  orig: c1
  dest: s2
  up: 100Mbps
  down: 100Mbps
  latency: 10
  time: 210 
 action: leave
  name: sv
  time: 240
\end{lstlisting}
\end{minipage}
\end{figure}

The \textbf{Emulation Manager} (EM) is the main component of \sys. \added[id=SI]{An instance of EM is deployed on every physical server involved in the experiment. Each instance is 
responsible for  providing containers collocated on the same physical host with their emulated end-to-end network properties.} 

Since \sys does not directly emulate network elements nor their internal state, we must accurately describe the topology at the end hosts.
This is achieved as follows.
\replaced[id=SI]{The EM starts}{We start} by parsing the topology description (\emph{e.g.}, Listing~\ref{kol:yaml:static}) into a graph structure, maintained throughout the emulation.
Next, \replaced[id=SI]{the EM computes}{we compute} the shortest paths between every pair of reachable containers.
Each shortest path is composed of several links, whose properties are used to determine the end-to-end network properties.
Formally, given a path \(\mathcal{P}\) composed of links \(\mathcal{P} = \{l_1, l_2, \ldots, l_n\}\), the end-to-end properties of the path can be computed as follows:

\begin{center}
$Latency(\mathcal{P}) = \sum_{i=1}^{n} Latency(l_i)$

$Jitter(\mathcal{P}) = \sqrt{\sum_{i=1}^{n} Jitter{(l_i)}^2}$

$Loss(\mathcal{P}) = 1.0 - \prod_{i=1}^{n} (1.0 - Loss(l_i))$

$\max{Bandwidth(\mathcal{P})} = \min_{\forall l_i \in \mathcal{P}} Bandwidth(l_i)$
\end{center}


For latency, packet loss and jitter (assuming a uniform distribution), it is enough to sum or multiply the properties of the links (variance for the jitter case). 

Bandwidth requires more considerations though, because it is limited not only by the physical capacity of the path, but also by all active flows on each link.
The maximum bandwidth in the path is determined by the link with the least bandwidth.
However, the bandwidth allocated to each active flow depends on all active flows in the same path and thus it must be dynamically recomputed at runtime.
Moreover, when the bandwidth required by each flow surpasses the maximum available bandwidth, the links become congested and therefore we need a mechanism to ensure a fair allocation of bandwidth among the competing flows.
In a real deployment, when competing flows require more bandwidth than the available capacity, network elements such as routers and switches buffer packets to accommodate the excess load up to a point where the buffers overflow and packets are dropped.
Unreliable transport protocols (\emph{i.e.}, UDP) ignore packet loss but reliable transport protocols (\emph{i.e.}, TCP) have congestion control mechanisms to adjust the throughput with the goal of allowing all competing flows to get a fair bandwidth share.
In \sys, rather than modeling the internals of network elements, which is expensive, we rely instead on a model to compute a fair share of the bandwidth available for each competing flow.
In particular, we leverage the RTT-Aware Min-Max model~\cite{Kelly,Massouli2002}, which gives a share to each flow that is proportional to its round-trip time and is inspired by TCP Reno~\cite{padhye2000modeling}, a widely adopted congestion control implementation.

Formally, the fair share of a long-lived flow $f$ is given by:

\begin{center}
$Share(f) = \left(RTT(f) \sum_{i=1}^{n} \frac{1}{RTT(f_i)}\right)^{-1}$
\end{center}

where \(f \in \{f_1, f_2, \dots, f_n\}\) are active flows on a link.


This bandwidth sharing model gives the percentage of the maximum bandwidth any flow is allowed to use at capacity.
However, it does not guarantee that the available bandwidth on a link will be fully utilized, for instance when a given flow is not consuming all its available share.
Therefore, when the sum of shares of all active flows is less than the maximum bandwidth on the link, \replaced[id=SI]{the EM performs}{we perform} a maximization step that increases the share of the other flows, proportionally to their original shares. \added[id=MM]{Note that \sys enforces bandwidth sharing per destination, not per flow.}

\emph{Congestion.}
The chosen model computes, per each flow, the maximum available bandwidth allowed at any given time.
While this works well when bandwidth usage is below or at capacity, it produces unrealistic results when the cumulative bandwidth required by flows surpasses the maximum bandwidth capacity.
The reason for this is a complex interplay between the Linux kernel Traffic Control's shaping action, the congestion algorithms of reliable transport protocols, and the implementation of such transport protocols in the  kernel itself.
In a real deployment with TCP, the protocol throttles its throughput dynamically by observing reported packet loss or delay when the buffers at network devices overflow.
Unlike TCP, UDP is insensitive to packet loss and simply continues to send packets at the application sending rate.
A first approach to model packet loss due to congestion would be to dimension the buffers (queue sizes) in the TCAL following known network buffer sizing strategies~\cite{Vu-Brugier:2007:CRP:1198255.1198262,vishwanath2009perspectives}.
However, this approach does not work due to the differences in behavior between \tc{} queues and those found in a switch or router.
On one hand, when a buffer in a router or switch fills up, it drops further incoming packets.\footnote{This is a simplification, as in practice packets already in the queue might be dropped to make room for incoming traffic with higher priority.} 
On the other hand, when the \texttt{htb qdisc} queue (used by the kernel) is full, rather than dropping packets, it back-pressures the application. 
The reasons for these discrepancies between \tc{} and router queues are because modeling packet loss is done in \texttt{netem}~\cite{linuxnetem} and not in \texttt{htb qdiscs}, and also due to Linux's TCP Small Queues (TSQ)~\cite{tsq1} (since kernel v3.6). 
TSQ reduces the number of packets in \texttt{qdiscs} and device queues with the goal of reducing the RTT, hence mitigating buffer-bloat. 
It works by tracking the amount of data waiting to be transmitted, and when this surpasses a given limit, the socket is throttled down, preventing further packets to be enqueued. 
The impact of this depends on the application --- when writing to a socket, an application using blocking I/O would block, while an application using non-blocking I/O would observe zero bytes written.
From our perspective, this means that there is no packet loss upon congestion and thus congestion control algorithms sensitive to loss would behave differently when emulated in \sys{} than in practice.
We address this limitation as follows.
When we observe that the overall requested bandwidth surpasses the available bandwidth we leverage \texttt{netem} to drop packets per flow proportionally to the oversubscribed capacity.
This dynamic adjustment of packet loss according to the excess bandwidth exposes packet loss to the congestion avoidance algorithm, allowing TCP connections to adjust their throughput.

For each container in the local machine, the Emulation Manager spawns an \emph{Emulation Core} process attached to the network namespace of the respective containers.
Each Emulation Core is responsible for obtaining the container bandwidth usage, exchange this metadata with the neighboring Emulation Cores to update the emulation model described above, and enforce the topology constraints through the TCAL, described below.
This design has several advantages.
First, it allows \sys to support any containerized (including third-party, legacy, code-obfuscated) applications without modifications.
Second, Emulation Cores on the same physical machine exchange emulation metadata via shared memory, reducing computational and networking overhead.
Finally, it allows the Emulation Manager to aggregate the data from the local Emulation Cores and exchange it directly with the Emulation Managers on the other physical machines.
\added[id=VS]{Note that having an EM per physical host allows metadata traffic to scale with the number of physical hosts rather than the number of application containers. 
Further, each EM only computes the part of the topology that affects the local containers thus reducing the computational overhead.} 


\textit{Dynamic Topologies.}
The Emulation Core also enforces the dynamic features of the topology.
The \texttt{dynamic} topology elements are reflected by modifications to the graph structure discussed above. 
Rather than computing modifications to the graph on the fly while the experiment executes, we pre-compute offline all the modifications before the experiment starts, as an ordered sequence of graphs.
We resort to this approach because while computing all the required metadata (\emph{e.g.}, all-pairs shortest paths, end-to-end properties, etc.) is fast for small graphs (\emph{e.g.}, few milliseconds), for large graphs with thousands of nodes it could take several seconds thus precluding accurate emulation of sub-second dynamics.

The \textbf{TC Abstraction Layer (TCAL)}
interfaces with \sloppy{Linux's} Traffic Control (\tc)\replaced[id=SI]{. \tc{} is a Linux user-space tool that allows manipulating the}{ to set up the initial} network properties (\emph{i.e.}, latency, bandwidth, packet loss, jitter) and retrieving bandwidth usage \replaced[id=SI]{of active connections.}{, and modifying the maximum available bandwidth on paths.}
\added[id=SI]{To this end, the \tc{} exposes an interface that allows for applying filters to classify data and then manipulate the network properties of each class independently. 
To control the network properties for each class, \tc{} supports a wide range of  queuing disciplines ({\texttt qdiscs}) where the packets are enqueued before being sent. 
\sys leverages two different types of \texttt qdiscs:\emph{(i)}~hierarchical token bucket ({\texttt htb qdisc}), a type of qdisc to control the bandwidth of outgoing packets and \emph{(ii)}~\texttt netem qdisc, that allows to apply delay and packet loss. For each destination, \sys creates a {\texttt htb qdisc} that enforces the bandwidth allocated to flows towards that destination. Besides,  a {\texttt netem qdisc} is also used to apply latency, jitter, and packet loss.}

\deleted[id=SI]{For each destination, we create a hierarchical token bucket ({\texttt htb qdisc}) that enforces the bandwidth throttling.
Besides, we attach to each {\texttt htb qdisc} a {\texttt netem qdisc} used to apply latency, jitter, and packet loss.}

Outbound traffic is matched to {\texttt netem qdiscs} through an {\texttt u32} universal 32bit~\cite{u32} traffic control filter. 
The filter is a two-level hashtable that matches against the destination IP address of packets and directs them to their corresponding {\texttt netem qdisc}.
This two-level design is due to limitations in the {\texttt u32}, which does not provide a real hashing mechanism (for speed reasons) but just a simple index in a 256 position array.
With a \texttt{/16} netmask this could result in several collisions, degrading performance.
We map the third octet of the IP address to the first level and the fourth octet to the second level of the hashtable, achieving constant lookup times.
Traffic directed to the {\texttt netem qdisc} will first be subjected to the {\texttt netem} rules to enforce latency, jitter and packet loss.
When packets are dequeued from {\texttt netem}, they are immediately queued in the parent {\texttt htb qdisc}, to enforce bandwidth restrictions. 
The TCAL structures are queried and updated very frequently during each experiment, namely to retrieve bandwidth usage and enforce the dynamic properties during runtime. 
To minimize the overhead of these calls, we rely on {\texttt netlink} sockets~\cite{rfc3549} that communicate directly with the kernel, circumventing the need to periodically spawn a new \texttt{tc} process.

The \textbf{Dashboard} allows 
users to monitor the progress of their experiments via a graphical web-based interface (not shown). 
This dashboard shows a graph-based representation of the emulated topology, the status of the services, ongoing traffic and dynamic events. 

As shown in \S\ref{sec:evaluation}, the decentralized design and simplified emulation model allows \sys{} to achieve accuracy on par with state-of-the-art approaches, while scaling to thousands of containers. 

\section{Implementation}
\label{sec:implementation}
\sys components are implemented in Python (v3.6), C and C++, and available at \url{https://angainor.science/kollaps}.
It requires a Docker daemon (v1.12) running on each machine.
The Deployment Generator currently supports Docker Swarm (v1.12) and \ks (v1.14) by generating Docker Compose or Kubernetes Manifest files, respectively. 
Users can customize these files (as required by many real applications~\cite{burns2016borg}) before starting an actual deployment.
\textbf{Privileged bootstrapping.} In order for an application running inside a Docker container to use {\texttt tc} (as the TCAL does), it must be executed with {\texttt CAP\_NET\_ADMIN} capability~\cite{docker:cap}.
Although Docker allows executing applications in standalone containers with user-specified capabilities, this feature is currently unavailable for Docker Swarm.
We circumvent this limitation as follows.
We deploy a bootstrapping container (the \emph{bootstrapper}) on every Swarm node. 
Its job is to launch, on that machine and outside Swarm itself, the Emulation Manager (\textsc{Em}). 
The \textsc{Em} shares the PID namespace with the host and has elevated privileges. 
It has access to the local Docker daemon and monitors the local creation of new containers. 
Upon the creation of a new container, the \textsc{Em} launches an Emulation Core responsible for that container, as discussed in \S\ref{sec:system}.
We distinguish between containers whose network should be emulated by \sys and regular containers through a tag injected in the configuration by the Deployment Generator.
We expect future releases of Docker Swarm to allow for a simplified mechanism. 
When using \ks, such restrictions do not hold and the \textsc{Em} is indeed deployed automatically. 


\textbf{Integration with container orchestrators.}
The design of \sys facilitates the integration with existing Docker images and container orchestration tools (\emph{e.g.}, Docker Swarm, \ks). 
In addition to producing ready-to-deploy Compose/Manifest files, the \sys deployment toolchain must configure three key resources managed in very different manners by the mentioned container orchestrators: 
\emph{(1)} access to the orchestrator APIs, used at runtime for name resolution, 
\emph{(2)} the topology descriptor file, read by each \textsc{Em} instance to setup the initial network state and compute the graph of the dynamic changes, 
and \emph{(3)} the setup of multiple virtual networks.

\subsection{Emulation Core and TCAL}
\label{sec:emuCoreImpl}

The TCAL library provides an interface to setup the initial networking configuration, retrieve bandwidth usage, and modify the maximum available bandwidth on paths.
It is implemented in C for performance reasons and consists of 2693 SLOC. 
The Emulation Core is implemented in Python and consist of 2963 SLOC.

The execution is split into two stages: initialization and emulation loop.
Once the initial graph representation is built, this component resolves the names of all services to obtain their IP addresses via the internal Swarm discovering service or \ks's API.
Then, it runs an all-pairs shortest path graph traversal~\cite{dijkstra1959note} between the local instance and all the other reachable instances. 
Finally, it computes the properties of the collapsed topology as described previously.
The properties of the paths are then set up by the TCAL, before moving to the emulation loop stage.
The emulation loop maintains a data structure with the bandwidth usage of each \replaced[id=MM]{flow}{instance}.
It works by periodically executing the following steps:
\begin{inparaenum}[\it (1)]
\item clear the state of all local active flows;
\item obtain the bandwidth usage by querying the TCAL; 
\item disseminate the local bandwidth usage to the other instances;
\item compute bandwidth usage on each path and its constituent links;
\item enforce bandwidth restrictions.
\end{inparaenum}
In parallel to the above algorithm, each Emulation Core instance collects the data sent at step (3) by the other Emulation Core instances.
This is used at (4) to compute the global bandwidth usage on a path and link basis.

\subsection{Metadata dissemination} 
\label{sec:needComms} 
All metadata is disseminated via Aeron~\cite{aeron}, an open-source, efficient and reliable UDP and IPC message transport protocol. 
For containers on the same machine, the metadata is exchanged through shared memory. 
This is possible because all Emulation Core processes are running on the Emulation Manager's process namespace. 
Across remote machines, metadata is disseminated through UDP messages. 
For each Emulation Manager there is an \emph{Aeron Media Driver} responsible for the dissemination of messages.
Every process reads from and writes to the Media Driver using a C++ library. 
The metadata messages embed the following fields:
\begin{inparaenum}[\it (i)]
\item number of flows, 2 bytes; 
\item list of used bandwidth per flow, 4 bytes per flow;
\item number of links;
\item list of link identifiers.
\end{inparaenum}
For emulated networks with $\leqslant$ 256 nodes, it is possible to pack the metadata information for links and identifiers in a single byte each (2 bytes are used for bigger emulated topologies).
As shown in \S\ref{sec:evaluation}, this approach allows to fit into a single UDP datagram as much information as possible, reducing the metadata traffic.

\section{Evaluation}\label{sec:evaluation}
We evaluated \sys through a series of micro- and macro-benchmark experiments in our cluster.
Furthermore, to validate the soundness of our approach against realistic scenarios, we compare the behavior of applications running on Amazon EC2 and under \sys.
Overall, our results show that:
\begin{inparaenum}[\it (1)]
\item \sys scales with the number of flows and containers, and has constant cost regardless of the emulated bandwidth usage;
\item running an application with \sys in a cluster or in Amazon EC2 yields similar results; 
\item \sys emulation accuracy is comparable with, and in some scenarios better than, tools that emulate the full network state such as Mininet. 
\end{inparaenum}


\begin{table}[!t]
    \caption{Bandwidth shaping accuracy for several emulated link capacities on a simple point-to-point client-server topology.
    \label{table:microbench:bw}}
    \setlength{\tabcolsep}{3pt}
  \scriptsize
  \center
  \rowcolors{1}{gray!10}{gray!0}
  \begin{tabular}{r|c|c|c|c}
  \rowcolor{gray!45}
  \textbf{Link BW} & \textbf{\sys} & \textbf{Mininet} & \textbf{trickle (def.)} & \textbf{trickle (tuned)} \\
    \hline
  \rowcolor{gray!25}
    Low (Kb/s) & & & &\\
    128 Kb/s & 122 (-5\%) & 123 (-4\%) & 262 (+104\%) & 131 (+2\%) \\
    256 Kb/s & 245 (-5\%) & 286 (+11\%) & 472 (+184\%) & 262 (+2\%) \\
    512 Kb/s & 490 (-5\%) & 490 (-5\%) & 717 (+40\%) & 525 (+2\%) \\
    \hline
  \rowcolor{gray!25}
    Mid (Mb/s) & & & &\\
    128 Mb/s & 122 (-5\%) & 122 (-5\%) & 250 (+95\%) & 131 (+2\%) \\
    256 Mb/s & 245 (-5\%) & 245 (-5\%) & 493 (-4\%) & 261 (+1\%) \\
    512 Mb/s & 487 (-5\%) & 486 (-5\%) & 952 (+85\%) & 518 (+1\%) \\
    \hline
  \rowcolor{gray!25}
    High (Gb/s) & & & &\\
    1 Gb/s  & 954 (-4\%) & 933 (-7\%) & 1.67  (+67\%) & 1.00 Gb/s \\
    2 Gb/s  & 1.91 (-4\%) & N/A & 1.93 (-3\%) & 1.97 (-1.5\%) \\
    4 Gb/s  & 3.79 (-7\%) & N/A & 4.12 (+3\%) & 3.61 (-10\%) \\
    \hline
  \end{tabular}
\end{table}

\textbf{Evaluation settings.} Our cluster is composed of 5 Dell PowerEdge R630 server machines, with 64-cores Intel Xeon E5-2683v4 clocked at 2.10~GHz CPU, 128~GB of RAM and connected by a Dell S6010-ON 40~GbE switch.
The nodes run Ubuntu Linux 18.04~LTS, kernel v4.15.0-65-generic.
The tests conducted on Amazon EC2 use \texttt{r4.16xlarge} instances, the closest type in terms of hardware-specs to the machines in our cluster.
We use the latest stable releases of Mininet (v2.2.2) and Maxinet (v1.2).
\subsection{Link-level Emulation Capabilities}
\label{eval:linklevelcap}
First we evaluate the accuracy of our bandwidth shaping mechanism under a topology that consists of two services running iPerf3~\cite{iperf}, connected by a single link.
\added[id=SI]{iPerf3 is a tool that measures the maximum bandwidth between its client and server instances. In this experiment, we use iPerf3 to assess the accuracy of \sys to emulate} a range of different target bandwidths, and 
compare the results \replaced[id=SI]{with}{with the same experiment executed with} Mininet~\cite{lantz2010network} and Trickle~\cite{eriksen2005trickle}, a userspace bandwidth 
shaper.
\replaced[id=SI]{During the experiment, we run iPerf3  for 60 seconds, and report the average bandwidth}{The experiment runs for 60 seconds and the average  is} in Table~\ref{table:microbench:bw}.
The values obtained with \sys and Mininet are similar since both systems rely on the \texttt{htb qdisc} to perform the bandwidth shaping.
Mininet however does not allow imposing bandwidth limits greater than 1Gb/s.
\sys does not impose that restriction and ensures the same level of accuracy of both systems at lower bandwidth rates ($\approx 95\%$).
Results using the default Trickle settings deviate significantly from the specified \replaced[id=SI]{bandwidth}{throughput} rates.
After a more detailed investigation, we were forced to tune iPerf3 to use smaller TCP sending buffers to achieve accuracy comparable with the other systems.

Next, we evaluate the accuracy of jitter emulation. 
For this, we set up a sequence of experiments \replaced[id=SI]{using the same topology of two nodes connected through a single link}{consisting of a topology with 2 services}, with one sending 10,000 \texttt{ping} requests to the other.
\replaced[id=SI]{We assign the link different latency values according to the measured latencies between services deployed on \texttt{us-east-1} and other Amazon AWS regions}{
We compare against collected average latency and jitter values between similar services deployed on \texttt{us-east-1} and various Amazon AWS regions}. 
Table \ref{table:microbench:jitter} shows for each destination AWS region (2\textsuperscript{nd} column), the measured latency and jitter values in the 3\textsuperscript{rd} and 4\textsuperscript{th} columns, respectively. On the right-most column, we present the jitter value emulated by \sys using the same latency.
The overall mean squared error between the \replaced[id=SI]{observed and emulated}{expected and observed} jitter is 0.2029.
While smaller errors could be achieved by directly controlling the network infrastructure on which \sys is deployed, this is beyond the scope of this work.

\begin{table}[!t]
    \caption{Jitter shaping accuracy for several emulated links with source at \texttt{us-east-1}.\label{table:microbench:jitter}}
	\centering
	\setlength{\tabcolsep}{3pt}
	\scriptsize
	\center
	\rowcolors{1}{gray!10}{gray!0}
	\begin{tabular}{l|c|c|c}
		\rowcolor{gray!25}
		\textbf{$\rightarrow$ Destination} & \textbf{Latency \added[id=SI]{(ms)}} & \textbf{EC2 Jitter \added[id=SI]{(ms)}} & \textbf{\sys Jitter \added[id=SI]{(ms)}} \\
		\hline
		us-east-1		&	6	&	0.5607	&	0.6367	\\	
		us-east-2		&	17	&	1.2411	&	1.4018	\\	
		ca-central-1	&	24	&	1.2451	&	1.3872	\\	
		us-west-1		&	70	&	1.3627	&	1.5438	\\	
		eu-west-1		&	78	&	1.2000	&	1.3684	\\	
		eu-west-2		&	85	&	1.6609	&	1.8592	\\	
		eu-north-1		&	119	&	1.2850	&	1.4479	\\	
		ap-northeast-1	&	170	&	1.4217	&	1.6031	\\	
		ap-south-1		&	194	&	2.0233	&	2.2758	\\	
		ap-northeast-2	&	200	&	1.8364	&	2.0888	\\	
		ap-southeast-2	&	208	&	1.4277	&	1.6290	\\	
		ap-southeast-1	&	249	&	1.2111	&	1.3728	\\	
		\hline
	\end{tabular}

\end{table}


\subsection{\replaced[id=SI]{Scalability}{Overhead of Metadata Traffic}}
\label{sec:evalMicro}
\sys relies on metadata dissemination to model and emulate bandwidth restrictions for competing flows.
\added[id=SI]{In this section, we assess the scalability of \sys by analysing metadata traffic growth and emulation accuracy with an increasing number of physical nodes.}

\begin{figure}[!t]
\includegraphics[width=1.0\linewidth]{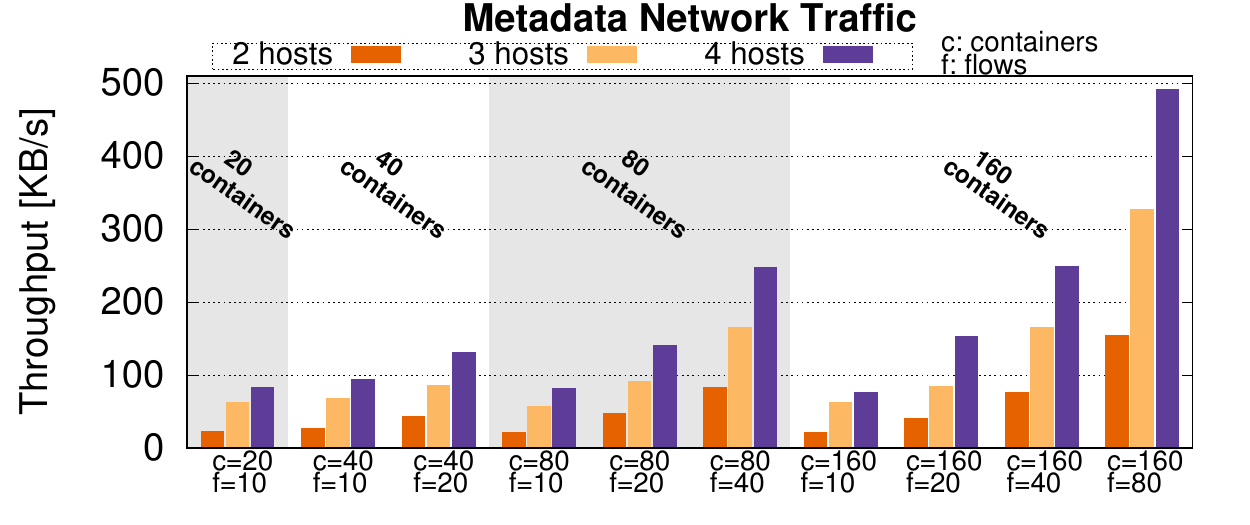}
\caption{\sys metadata \replaced[id=VS]{network usage}{bandwidth} with an increasing number of containers (C)\replaced[id=MM]{, }{ and} flows (F)\replaced[id=MM]{, and hosts}{ when deployed across 2, 3 and 4 physical hosts}.}
\label{plot:metadata}
\vspace{-8pt}
\end{figure}

\replaced[id=SI]{First}{In this section}, we study the cost of metadata dissemination by deploying several dumbbell topologies\deleted[id=SI]{ similar to the one in Figure \ref{fig:topology}}, across a cluster of increasing size. \added[id=SI]{The containers are distributed evenly among the physical nodes, with the client containers on one side and the server containers on the other side of the dumbbell topology.}
We use iPerf3 to generate steady TCP traffic of $50 Mb/s$, the maximum capacity of the shared link. 
We denote each configuration by a tuple \replaced[id=SI]{\texttt{(C,F)}}{\texttt{C/F}}, with \texttt{C} the total deployed containers and \texttt{F} the number of concurrent flows. 
Results with clusters of up to four physical machines are shown in Figure~\ref{plot:metadata}. 

As discussed in \S\ref{sec:system} and \S\ref{sec:implementation}, \replaced[id=SI]{\sys}{the Emulation Manager} uses shared memory \replaced[id=SI]{on top of}{and} the network (through Aeron) to exchange metadata among the local Emulation Cores and the Emulation Manager components residing on \replaced[id=SI]{each}{other} machine, respectively.
\replaced[id=SI]{As expected}{Hence}, running on one single machine leads to zero bandwidth usage, \added[id=SI]{and increasing the number of physical hosts increases the metadata 
bandwidth. Interestingly, though, }
we observe that the metadata traffic is not affected by the number of containers. \added[id=SI]{This is because: \emph{(i)} only active flows require the exchange of metadata and \emph{(ii)} bandwidth sharing is enforced per destination and not per flow, thus reducing the overall bandwidth.} \deleted[id=SI]{, as expected, while it does increase with the number of physical hosts.}

Note that metadata traffic does not increase with the emulated application bandwidth, since the messages have a constant size.
Overall, we observe that metadata traffic requirements are quite modest even for topologies with 160 containers deployed across 4 physical machines, \emph{i.e.}, $493 KB/s$.  
\added[id=SI]{Despite growing linearly with the number of physical hosts, the overall metadata traffic is negligible when considering the available bandwidth of the dedicated clusters we target with \sys. 
We further study this in the next experiment.
}

\begin{figure}[!t]
\includegraphics[width=1.0\linewidth]{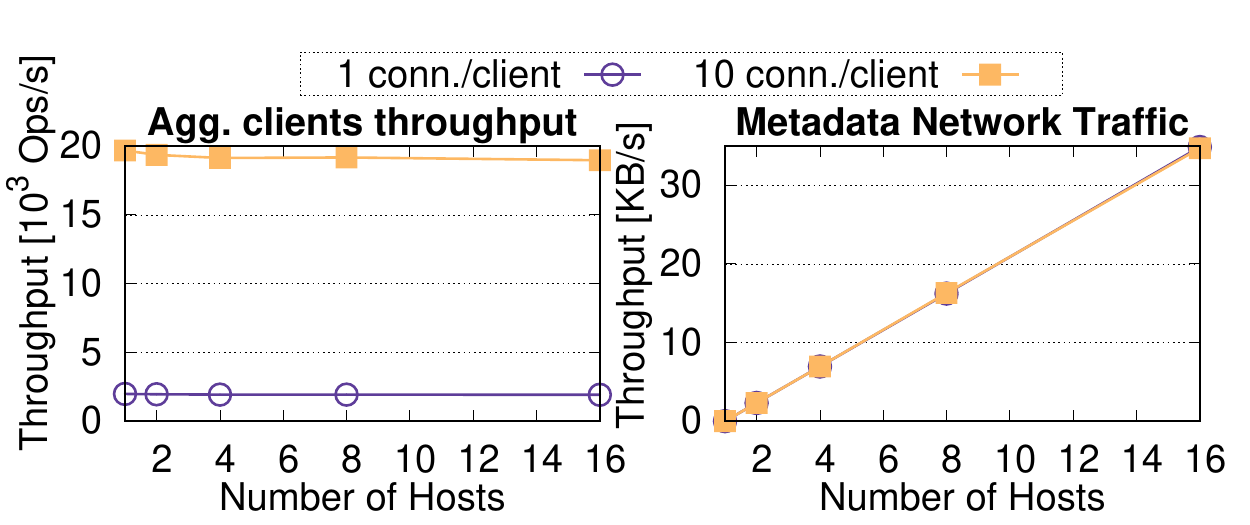}
\caption{\added[id=VS]{Aggregate throughput of twelve memcached clients (left) and the metadata traffic per host (right) for emulation on 1, 2, 4, 8 and 16 physical hosts. Note that the lines on the right figure overlap.}}
\label{plot:memcached}
\vspace{-8pt}
\end{figure}

\added[id=SI]{To study how metadata grows with an increasing number of nodes, and how distribution affects the emulation, we design the following experiment. 
We deploy \texttt{memcached}~\cite{memcached}, an in-memory key-value store, in a geo-distributed topology with 4 emulated Amazon AWS regions~\cite{Sousa2016} (also used in the experiments of \S\ref{sec:evalReal}).
We place a \texttt{memcached} server container on each region, collocated with three client containers. 
Each server handles two local clients and a remote one. 
Clients execute the \texttt{memtier} benchmark~\cite{memtier} for key-value stores. 
Figure~\ref{plot:memcached} reports the results for two different configurations of the client, with 1 and 10 connections per client, which correspond to an increase in the number of the flows. 
The target load was selected to fit the experiment in a single physical machine thus allowing us to observe how the distribution across an increasing number of physical machines affects the application metrics.
We deploy the experiment on top of 1, 2, 4, 8 and 16 physical machines.}

\added[id=SI]{We report the aggregate throughput of all the twelve clients and the metadata traffic consumed per physical hosts. 
The aggregate throughput (Figure~\ref{plot:memcached}, left) for both configurations (1 and 10 connections per clients) is consistent across the number of hosts.
This results confirms that the decentralized nature of \sys allows to emulate the same behavior with a larger number of physical hosts.
Looking at the metadata traffic (Figure~\ref{plot:memcached}, right), it indeed increases with the number of physical hosts but the overall value is negligible when compared to the available bandwidth in the target cluster.
}

\begin{figure}[!t]
    \includegraphics[width=1.0\linewidth]{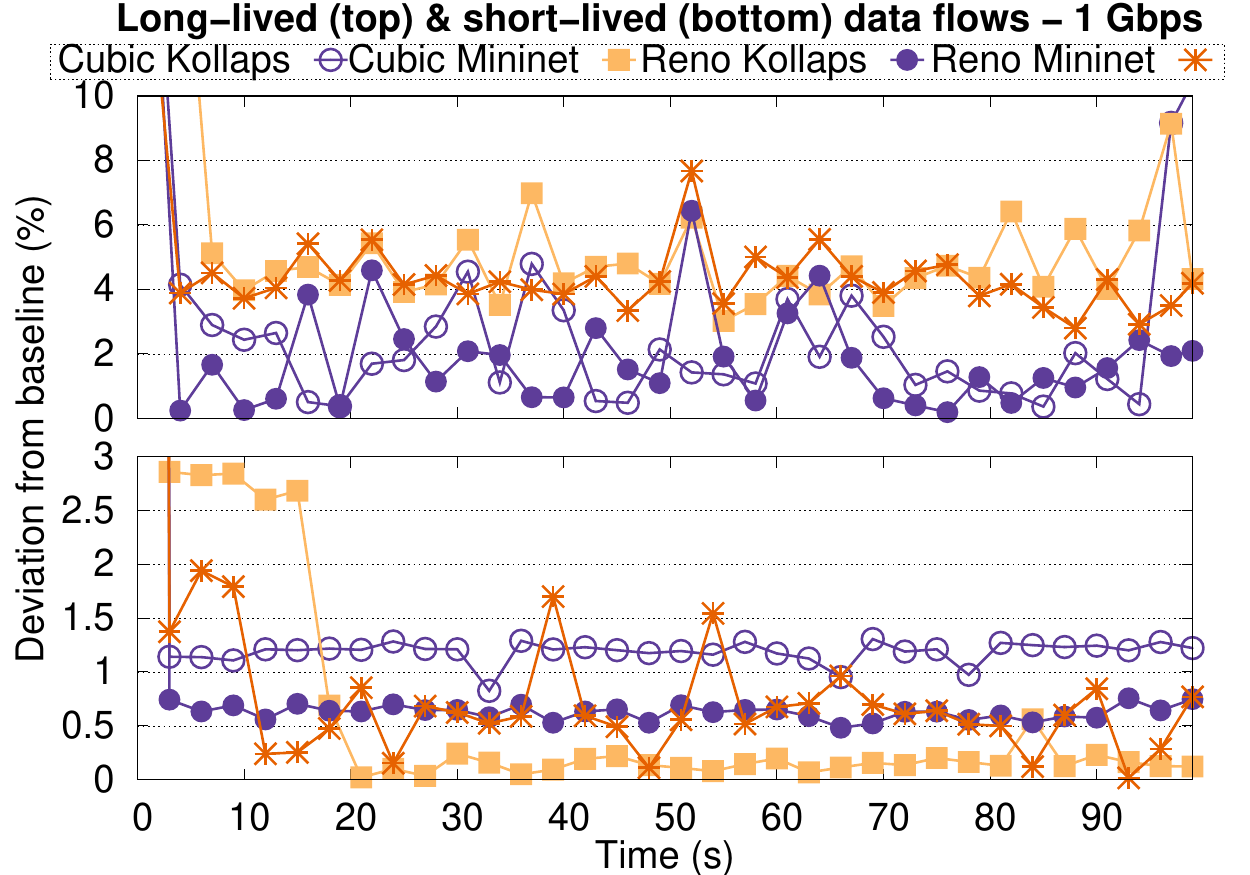}
    \caption{Error rate of bandwidth measures for long-lived (top) and short-lived (bottom) flows in \sys and Mininet when compared to a bare-metal deployment. Flows correspond to iPerf3 and wrk2 traffic respectively.
    \label{plot:long-short-lived-error-rate}}
    \vspace{-12pt}
\end{figure}

\subsection{Long- and short-lived flows}
We now study the accuracy of \sys when handling short-lived and long-lived flows.
To this end we set up a bare-metal experiment with one server and two clients interconnected through a 1Gb/s switch.
For the long-lived flows we use iPerf3 and \texttt{tcpdump} (v4.9.2) to measure the throughput of Cubic~\cite{Ha2008} and Reno~\cite{jacobson1988congestion} congestion control algorithms over 100 seconds.
We then repeat the experiment with Mininet and \sys and compute the observed deviation (error) from bare metal as $\left|1 - \frac{observed~bandwidth}{baremetal~bandwidth}\right|$.
Results are depicted in Figure~\ref{plot:long-short-lived-error-rate} (top).
Note that due to the unpredictable nature of congestion control algorithms the overall deviation is more important than \replaced[id=SI]{momentary}{pontual} fluctuations over time. 
Interestingly, we observe that in general \sys is closer to the bare metal observations than Mininet, despite the former fully modeling the full state of the network switch.

We repeat the same experiment for short-lived flows using \texttt{wrk2}~\cite{wrk2}, a popular HTTP benchmarking tool. 
This tool maintains a set of open connections (\emph{i.e.},  100 connections over 2 threads, the default configuration) and executes continuous HTTP requests ($\sim$64KB each) over them for a given duration. 
\deleted[id=SI]{Such workload injects short flows over the same connection.}
Results are depicted in Figure~\ref{plot:long-short-lived-error-rate} (bottom).
In  this scenario, \sys is on-par with Mininet, both with a relative error smaller than $2\%$ for most of the experiment.

\begin{figure}[!t]
    \includegraphics[width=1.0\linewidth,trim={0 20pt 0 10pt}]{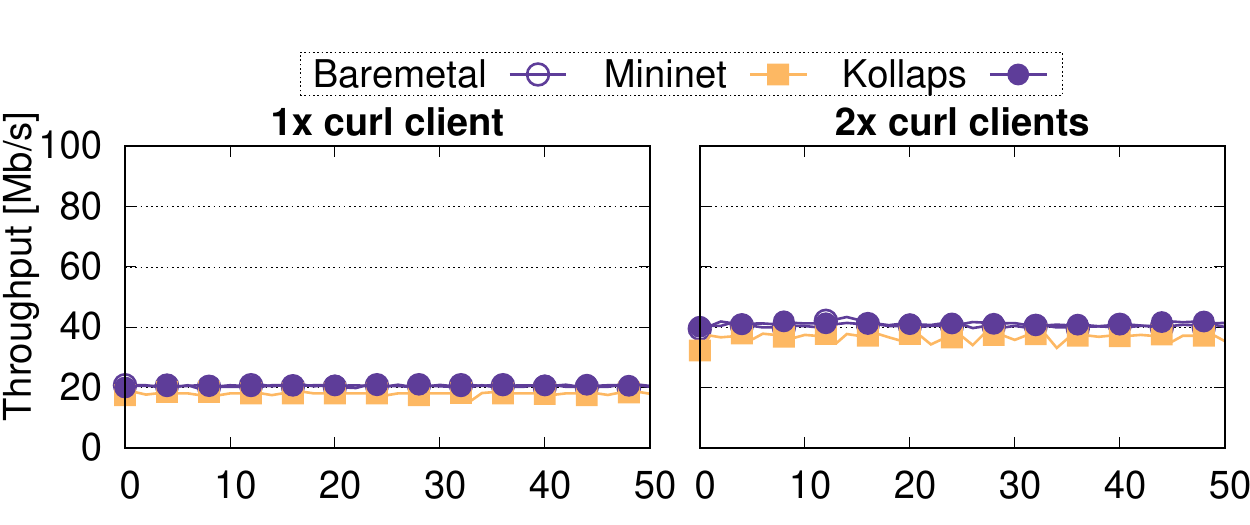}
    \\
    \includegraphics[width=1.0\linewidth,trim={0 0 0 10pt},clip]{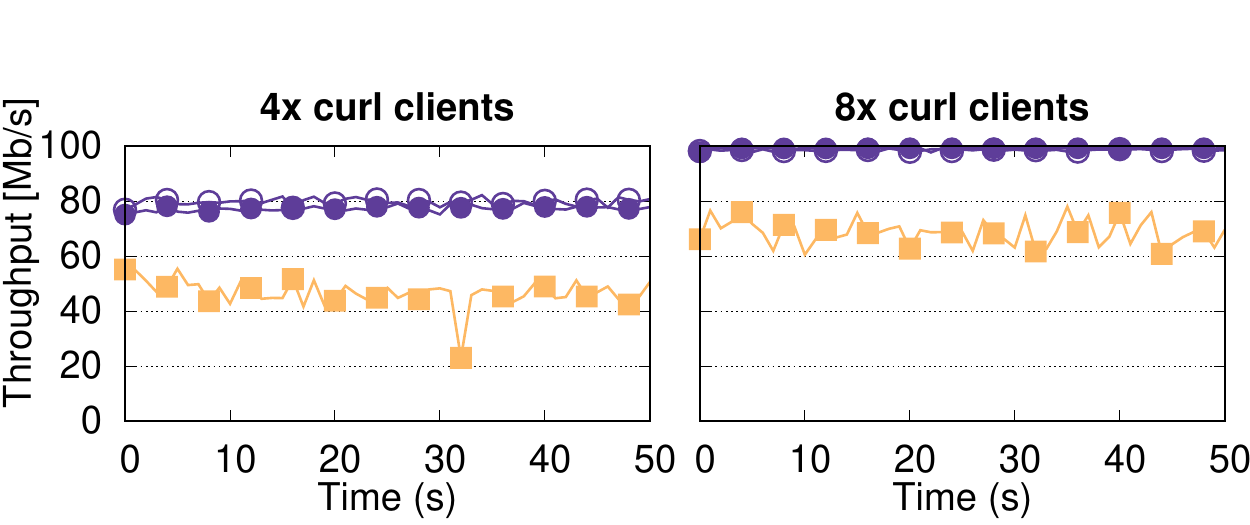}
    \caption{Throughput of a HTTP server serving different number of \texttt{curl} clients using \sys and Mininet when compared to a bare-metal deployment.
    \label{plot:http-sever}}
    \vspace{-12pt}
\end{figure}

Note that \texttt{wrk2} keeps a connection open and sends several small requests over the same connection.
We now show the results for small requests sent over short connections, \emph{i.e.}, each request starts a new TCP connection.
To this end, we use the same request size used for \texttt{wrk2}, but instead rely on \texttt{curl}~\cite{curl}, a popular command-line HTTP client. 
In this experiment we use a 100Mb/s bandwidth link. 
We varied the number of concurrent clients from 1 to 8 to control the load on the network. 
Figure~\ref{plot:http-sever} shows the result for this setting. 
As expected, increasing the number of clients increases the load proportionally, as each client is independent and the server has sufficient capacity (\emph{i.e.}, memory and cpu) to accomodate the increasing load.
We can see that \sys provides similar throughput to bare-metal deployment across  the different configurations. 
On the contrary, in such workload, Mininet fails to keep up \replaced[id=SI]{as the client load increases.}{with the load with the difference to bare-metal increasing with the  client load.}
We explain this behaviour by the fact that Mininet needs to maintain the full state of the switches, which becomes a significant overhead when the number of (short) connections grows.

\begin{figure}[!t]
    \includegraphics[width=\linewidth]{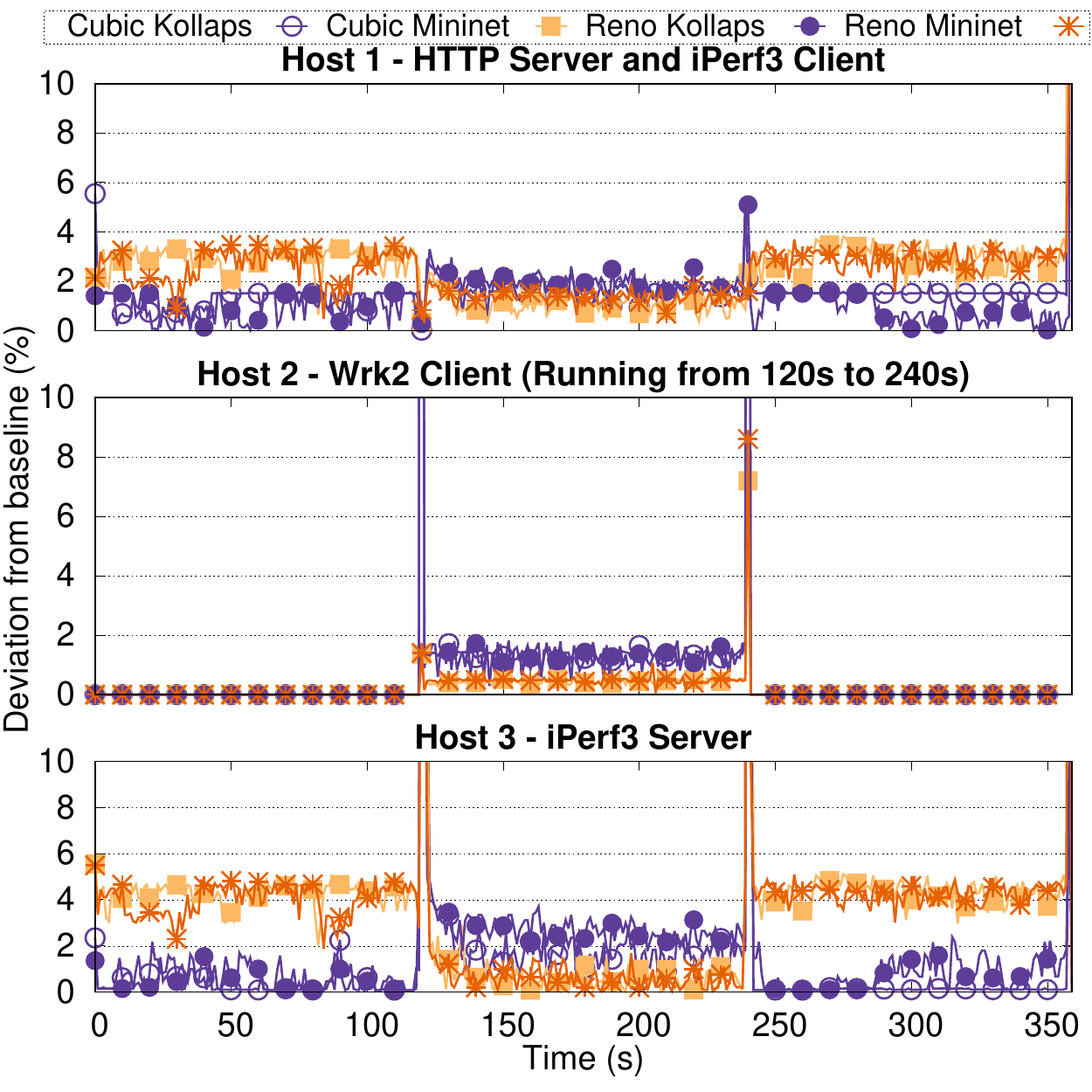}
    \caption{Error rate of bandwidth measurements in a mixed short- and long-lived flows deployment. We compare the measures recorded from host 1 (wrk2 server and iPerf3 client - top), host 2 (wrk2 client - middle) and host 3 (iPerf3 server) for \sys and Mininet against a bare-metal baseline.
    \label{plot:mixture-error-rate}}
\end{figure}

Lastly, we design an experiment involving both short- and long-lived data flows.
We set up three hosts: host one running an \texttt{HTTP} webserver and an iPerf3 client, host two running a wrk2 client querying host one, and host three running an iPerf3 server being queried by host one.
We run this setting for 6 minutes.
Until minute two, only the long-lived (iPerf3) client is sending data.
From minute two to minute four, the short lived (\texttt{wrk2}) client tries to utilize the link to its full capacity.
For the remainder, only the long-lived client \replaced[id=SI]{is sending data, like the first two minutes}{runs}.
Figure~\ref{plot:mixture-error-rate} presents the deviation from the bare-metal deployment, when reproducing this scenario with \sys and Mininet.
We observe that the error rate for Mininet and \sys are comparable and mostly below 5\% with only a spike in the transition period in both systems.

\subsection{Decentralized Bandwidth Throttling}
\label{sec:evaldecent}

Next we investigate the effectiveness of our bandwidth sharing model when the bandwidth requested by the application exceeds the available capacity.
To assess this, we set up a topology with six clients (\texttt{C1 - C6}), three bridges (\texttt{B1 - B3}) and 6 servers (\texttt{S1 - S6}). 
The first three clients are connected to \texttt{B1} through links with bandwidths of $50$, $50$ and $10Mb/s$ and latencies of $10$, $5$ and $5ms$, respectively. 
The other three clients are connected to \texttt{B2} with the same links properties.  
All servers are linked to \texttt{B3} through equal links with $50Mb/s$ bandwidth and $5ms$ latency. 
Finally, \texttt{B1} is connected to \texttt{B2} by a $50Mb/s$ link with $10ms$ latency, and \texttt{B2} is connected to \texttt{B3} by a $100Mb/s$ link with $10ms$ latency. 

%

\begin{figure}
	\includegraphics[width=1.0\linewidth]{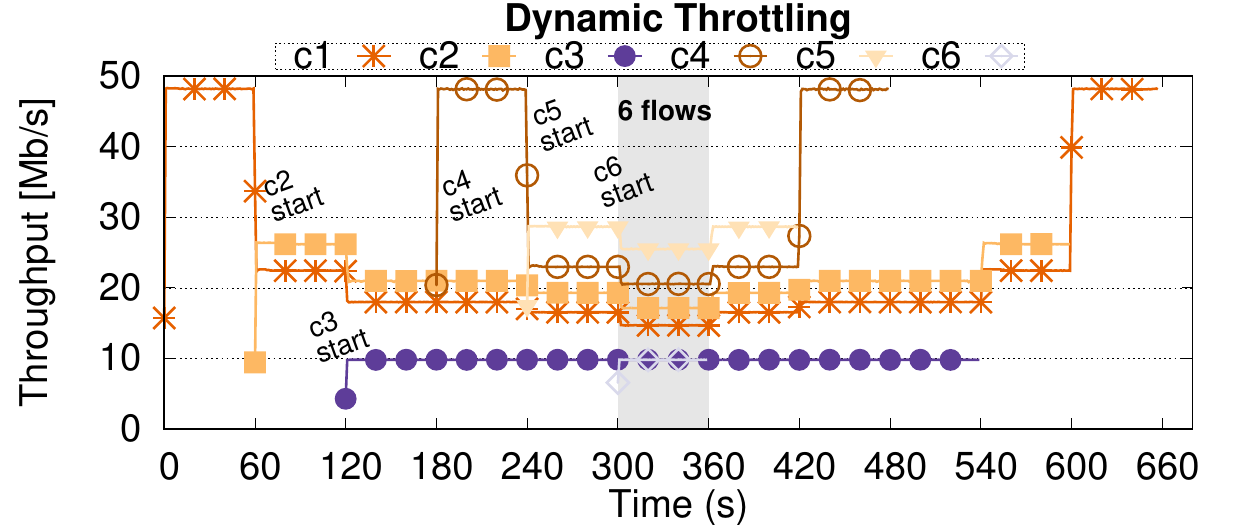}
	\caption{Decentralized bandwidth throttling: several clients compete on a shared link. Each gets a different share of bandwidth, adjusted at runtime.}
	\label{plot:decentralized-throttling}
\end{figure}

\added[id=SI]{Figure~\ref{plot:decentralized-throttling} shows the bandwidth of each of the 
established flows along the time}. We use iPerf3 to establish continuous TCP flows between clients and servers, while the experiment proceeds as follows.
In the first half of the experiment we start each client sequentially in $60$ second intervals. 
Initially, only \texttt{C1} has an active flow, and hence it uses all the available bandwidth. 
Upon starting \texttt{C2}, both clients will compete for the bandwidth over the shared links. 
At this point, since \texttt{C2} has a smaller RTT than \texttt{C1}, it gets a proportionally higher share of the bandwidth. 
Following the model in \S\ref{sec:system}, these shares are $23.08~Mb/s$ and $26.92~Mb/s$, respectively. 
When \texttt{C3} starts, it will be allowed an equal share of the bandwidth to \texttt{C2}. 
However, \texttt{C3} is limited by a $10Mb/s$ link prior to the contended one. 
Consequently, the bandwidth share of the other two clients is increased proportionally to their original shares, resulting in $18.45$, $21.55$, and $10~Mb/s$, respectively.

At $180$ seconds, \texttt{C4} starts. 
It can reach $50Mb/s$ because the throughput of all other three clients is limited by the $50~Mb/s$ link connecting the bridges \texttt{B1} and \texttt{B2}. 
Hence, the link between \texttt{B2} and \texttt{B3} can accommodate all four clients. 
When \texttt{C5} starts, this is no longer the case. 
Now, all five clients are competing for the $100~Mb/s$ link. 
\texttt{C3} remains limited to $10~Mb/s$, below its allowed share. 
The shares for all other clients is increased accordingly resulting in $16.89$, $19.75$, $10$, $23.74$, and $29.62~Mb/s$, respectively.
At $300$ seconds, \texttt{C6} starts and, like \texttt{C3}, the maximum bandwidth it can use is lower than its given share. 
The expected bandwidths therefore become $15.04$, $17.55$, $10$, $21.06$, $26.33$, and $10$ $Mb/s$ for clients \texttt{C1-C6}, respectively.

On the second half of the experiment (from $360s$ until the end) we sequentially shutdown the clients every $60s$ in the reverse order of arrival. 
Despite the decentralized emulation model, \sys is able to quickly adjust the bandwidth shares to the dynamic behavior of clients.

\subsection{Large-scale topologies}\label{sec:evalMini}

We now compare \sys with Mininet~\cite{lantz2010network} and Maxinet~\cite{Wette2014} in large-scale topologies.

We consider large-scale topologies generated using the preferential attachment algorithm~\cite{Barabasi1999}.
This method yields scale-free networks, which are representative of the characteristics of Internet topologies.
The experiment consists of end-nodes sending ICMP echo requests (\texttt{ping}) to other random end-nodes for 10 minutes.
We compare the obtained round-trip times (\texttt{RTT}) with the theoretical ones statically computed from the topology.
The results are presented in Table~\ref{table:scalefree} as a mean squared error between these two quantities for topologies with 1000, 2000, and 4000 elements.
\begin{table}[!t]
  \caption{Mean squared error exhibited on latency tests with large scale-free topologies in \sys, Mininet and Maxinet.
      }
    \setlength{\tabcolsep}{3pt}
    \scriptsize
    \center
    \rowcolors{1}{gray!10}{gray!0}
    \begin{tabular}{r|c|c|c|c|c}
        \rowcolor{gray!25}
  \textbf{Topology size} & \textbf{\#Nodes} & \textbf{\#Switches} & \textbf{\sys} & \textbf{Mininet} & \textbf{Maxinet}\\
    \hline
  1000 & 666 & 334 & 0.0261 & 0.0079 & 28.0779\\
  2000 & 1344 & 656 & 0.0384 & NA & 347.5303\\
  4000 & 2668 & 1332 & 0.0721 & NA & NA\\
      \hline 
\end{tabular}

\label{table:scalefree}
\end{table}
\sys and Maxinet are deployed on four machines while Mininet is deployed in a single machine as a multiple machine deployment is not supported.
We observe that Mininet produces smaller errors than \sys for the 1000 topology.
The reasons are twofold. 
First, the container networking in Docker introduces small yet measurable delays.
Second, because \sys is running on different physical machines, there is also a small but measurable delay when packets need to traverse the physical links.
Despite these two factors, the largest deviations from the theoretical \texttt{RTTs} observed with \sys were $0.27ms$, $0.4ms$ and $0.55ms$ for the 1000, 2000, and 4000 topologies, respectively. For reference, the minimum theoretical \texttt{RTTs} in the three topologies are $10ms$, $22ms$ and $14ms$, respectively. \added[id=SI]{Accordingly, the deviation values correspond to a MSE of 0.0261, 0.0384, and 0.0721, respectively, as depicted in the Table~\ref{table:scalefree}.}

Due to the current limitations with Mininet,
it was not possible to gather results for the larger topologies. 
Maxinet requires an external controller to manage the emulated switches.
We experimented with several configurations with POX~\cite{pox} modules, Floodlight~\cite{floodlight}, and Opendaylight~\cite{opendaylight} to find out which one yielded the best results.
The controller configuration used for these experiments rely on 4 distinct POX controllers executing the \texttt{forwarding.l2\_nx module}, the best performing one for this scenario.
The error obtained for Maxinet is significantly higher than both \sys and Mininet, with the largest deviation reaching 11ms and 40ms on the 1000 and 2000 topologies respectively, larger than the minimum theoretical delays in each topology.
We attribute this to the overhead of having an external controller, as well as to the type of controller (with other configurations producing even worse results).
For these reasons, we did not run further experiments for Maxinet in the 4000 topology.
To a lesser extent, Maxinet also suffers from the small yet measurable delay when packets need to traverse the physical links.



\subsection{Geo-replicated systems}
\label{sec:evalReal}
\begin{figure}[!t]
\includegraphics[width=1.0\linewidth]{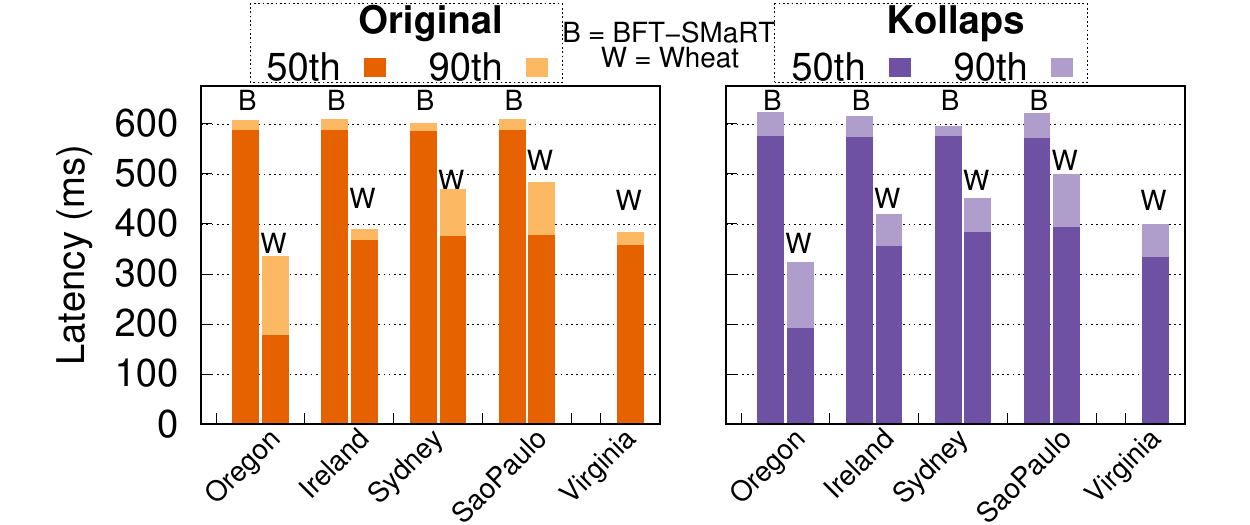}
\caption{Reproduction of an experiment with a geo-replicated deployment of BFT-SMaRt and Wheat. The experiment measures the latencies of clients located in different Amazon EC2 regions (left: results from~\cite{Sousa2016}, right: same experiments with \sys).}
\label{fig:bft}
\end{figure}

We turn our attention to macro-benchmarks to assess and motivate the behavior of \sys in real-world scenarios.

\textbf{Reproducibility.}
In this experiment, we reproduce results obtained for two Byzantine fault tolerant state machine replication libraries: BFT-SMaRt~\cite{Bessani2014a}, and its optimized version Wheat~\cite{Sousa2016}.
The authors of these systems evaluate and compare them through a geo-distributed deployment on Amazon EC2 instances spanning 5 regions~\cite{Sousa2016}.
The experiment consists of placing one server and one client at each region, with servers running a simple replicated counter.
Aside from the experimental results, the authors also provide the measured average latency and jitter between regions (~\cite{Sousa2016}, Table II), which we use to model a topology in \sys that mimics  the one observed in their experiments.

Figure~\ref{fig:bft} shows the results of the original experiment on EC2 (left), and using \sys (right). 
As we can  observe, the results of executing the experiment in \sys are close to the results achieved by the authors on EC2, with a maximum difference of 7.3\% observable between the 90th percentiles of the Wheat client in Ireland.
BFT-SMaRt results were even closer, with a maximum difference of 2.7\%.

We attribute the difference to the following.
Since the authors only provide the average and standard deviation latency measurements, we assumed for the \sys experiments a normal jitter distribution.
However, the Amazon EC2 \texttt{t1.micro} instances used by the authors in their experiments are prone to jittery behavior~\cite{EC2.t1micro}, potentially not following a normal distribution.
This is relevant in particular to Wheat as its more latency sensitive than BFT-SMaRt and thus more affected by the jitter distribution.


\begin{figure}
\includegraphics[width=1.0\linewidth]{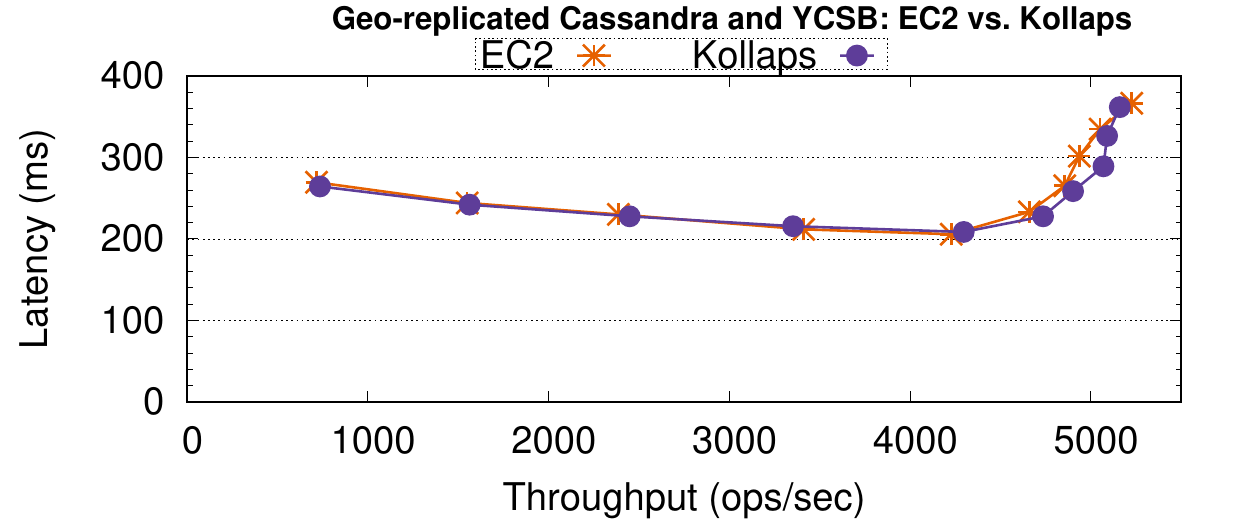}
\caption{Throughput/latency of a geo-replicated Cassandra deployment on Amazon EC2 and \sys.}\label{plot:cassandra}
\end{figure}
\textbf{NoSQL evaluation.}
We now compare the results of benchmarking a geo-replicated Apache Cassandra~\cite{cassandra,Lakshman:2010:CDS:1773912.1773922} deployment on Amazon EC2 and on \sys\added[id=SI]{ in our local dedicated cluster}.
The deployment consists of 4 replicas in Frankfurt, 4 replicas in Sydney and 4 YCSB~\cite{Cooper:2010:BCS:1807128.1807152} clients in Frankfurt.
Cassandra is set up to active replication with a replication factor of 2.
The YCSB operations are configured to require a quorum on updates and only one response on reads, with a 50/50 mix of reads and updates.
YCSB will direct most requests at the replicas in Frankfurt which are closer, however, a reply from the replicas in Sydney must always be present for a write quorum to succeed.
In order to model the network topology in \sys, we collected the average latency and overall jitter between all the Amazon EC2 instances used, prior to executing the experiment on Amazon.
Figure~\ref{plot:cassandra} shows the throughput-latency curve obtained from the benchmark on both the real deployment on Amazon EC2 and on \sys\added[id=SI]{ in our local  cluster}.
The results are a close match, showing only slight differences after the turning point where response latencies climb fast, as Cassandra replicas are under high stress.
This experiment illustrates how \sys can be used to assess the behavior of real systems in a controlled environment\added[id=SI]{ without requiring expensive real-life deployments}. 

\begin{figure}[!t]
\includegraphics[width=1.0\linewidth]{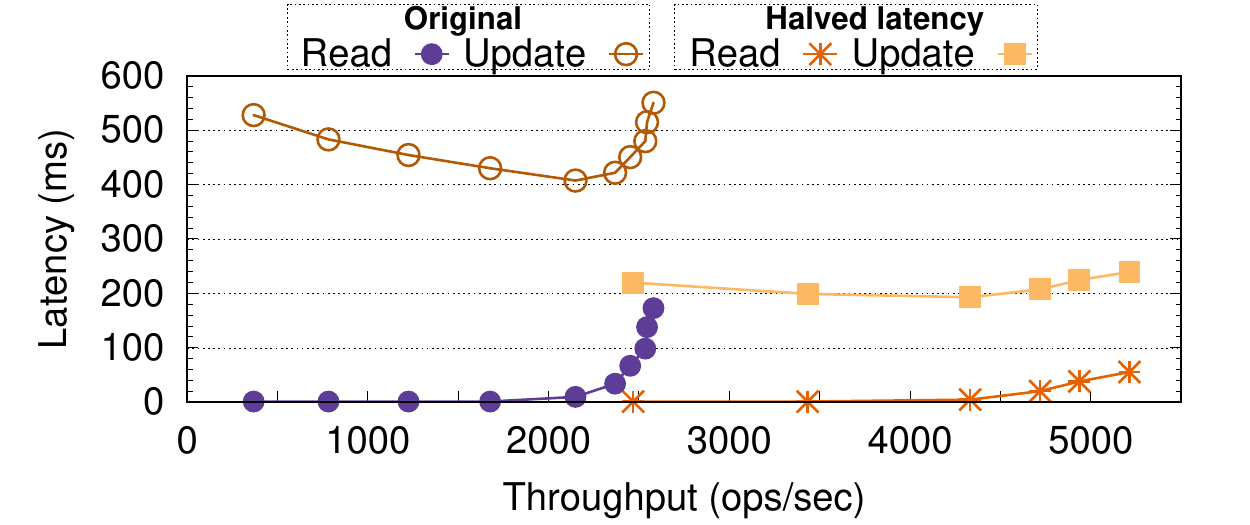}
\caption{Throughput/latency of a geo-replicated Cassandra on \sys using a hypothetical topology answering the question: \emph{what-if} nodes were moved from \texttt{ap-south} to \texttt{ap-northeast}?}\label{plot:cassandra_half}
\end{figure}

\textbf{The what-if use-cases.} Finally, we present a possible use-case for \sys, evaluating applications in an hypothetical \textit{what-if} scenario.
For instance, it might be useful to \replaced[id=SI]{study}{know} the behavior of Cassandra if the latencies between EC2 regions were to be halved.
In practice, this would correspond to the scenario of moving 4 Cassandra nodes from Sydney (\texttt{ap-south}) to Seoul (\texttt{ap-northeast}).
\added[id=SI]{Instead of relying on a costly and time-consuming real-life deployment, \sys enables this study with a simple change in the topology configuration file.} 
Figure~\ref{plot:cassandra_half} shows the obtained results.
For the sake of readability and to ease the comparison, we further split the previous results \added[id=SI]{(from the Frankfurt and Sydney deployment)} into read and write curves and show them alongside the hypothetical results  \added[id=SI]{(obtained from \sys emulation)} with the halved latency scenario.
In this case, Cassandra behaves as expected: request latencies drop by about half and Cassandra reaches higher throughputs.

\section{Limitations}
\label{sec:limitations}

We identify the following limitations in \sys.

\textbf{Interactivity.} 
For dynamic topologies, we compute off-line, and locally at each node, the sequence of all graph states over time.
While this approach allows to achieve sub-second emulation precision, it also prevents the support to establish an interactive testing session for which a precise crash plan is not defined statically by configuration but rather decided by the user on the fly.
In principle, it is possible to support interactive experiments by computing and applying the graph changes online at the expense of some accuracy.
The experimenter can afterwards decide, based on the reported error, if the experiment is satisfactory or if the same sequence of steps should be converted into a statically defined experiment. 

\textbf{Multipath routing.}
While the emulated topology itself can include multiple paths between each two pairs of nodes, \sys uses a shortest path algorithm to compute the collapsed links between every pair of containers, effectively discarding any multipath routing~\cite{cidon1999analysis} considerations.
We plan to support this in the future by: i) extending the language to allow the specification of multiple paths, ii) use a k-shortest paths algorithm for link collapsing, and iii) extend the emulation model to take this into account.

\textbf{Multicast.} 
Note also that \sys does not currently support multicast because the multicast tree is maintained at the network elements such as switches and bridges, which we do not model.

\textbf{Beyond the physical links.}
\sys only emulates network topologies whose aggregate capacity fits into the limits of the underlying physical cluster \added[id=SI]{(i.e., it is impossible to emulate a link of 10Gb/s if \sys is running on a cluster with 1Gb/s connections). 
Moreover, the fact that the bandwidth sharing is updated upon each iteration of the Emulation Manager forces a lower bound on the minimum latency that \sys can emulate.
For example, \sys will either fail to capture and update the bandwidth sharing for short flows that span a time interval shorter than a single iteration, or would react after the flow has ended. 
In this sense, our design and implementation is better suited for emulating WAN deployments rather than emulating data-center environments. 
Possible approaches to mitigate these limitations are discussed in Section~\ref{sec:conclusion}.}
\deleted[id=SI]{Crossing such limits requires time dilation~\cite{Gupta:2005:IBT:1095810.1118605} capabilities, both in \sys and on the containers themselves.
Recent work explores such support in containers in the context of SDN~\cite{Yan2017}, which we plan to leverage.}



\vspace{-10pt}
\section{Conclusion \added[id=SI]{and Future Work}} 
\label{sec:conclusion}
\vspace{-3pt}
The present work stems from the need to simplify the evaluation of large-scale geo-distributed applications.
Rather than emulating the full network state, we argue that application-level metrics are mostly affected by the macro network properties, such as end-to-end latency, bandwidth, packet loss and jitter.
We assessed the feasibility of this idea by designing, implementing and evaluating \sys a decentralized topology emulator.
Our experiments, on small and large-scale Internet-like topologies, in both static and dynamic settings, show that \sys is able to accurately reproduce real-world deployments of off-the-shelf popular systems, such as Cassandra.
To our community, reproducibility of results is increasingly important and we believe \sys can be a useful tool to achieve this goal.
We showed this by reproducing results from a geographically distributed state machine replication system presented in the literature~\cite{Sousa2016}.
Finally, \sys can also be used by engineers to predict application performance and correctness under hypothetical, but fully controlled, network conditions. 

\added[id=SI]{In future work, we plan to address several limitations of \sys. 
To emulate networks with capacities higher than the infrastructure upon which \sys is running,  
we plan to investigate time dilation~\cite{Gupta:2005:IBT:1095810.1118605} capabilities, both in \sys and on the containers themselves. 
This can also help in mitigating the limitation of flows shorter than a single iteration of the Emulation Manager, and therefore enable \sys to emulate data-center environments, as recently proven to help in the context of SDN emulation~\cite{Yan2017}.
To further enhance efficiency of the Emulation Manager and lowering the lower bound of the minimum latency that can be emulated, one can switch from periodic metadata dissemination to sending updates only upon changes in the flows. 
This approach reduces the metadata traffic, especially in the presence of long-lived flows, and will allow the Emulation Manager to react more promptly to short-lived flows.} 
\vspace{-3pt}

\begin{acks}
We are grateful to Allyson Bessani for having shared with us the datasets of BFT-Smart and Wheat.
We thank Jo\~ao Leit\~ao and Pedro Fouto for their help in deploying the Cassandra experiments.
We would also like to thank our shepherd KyoungSoo Park, and the anonymous referees for their valuable comments and helpful suggestions who led to this version of the paper. 
This work was partially supported by national funds through FCT, Fundação para a Ciência e a Tecnologia, under project UIDB/50021/2020 and
project Lisboa-01-0145-FEDER- 031456 (Angainor).


\end{acks}

	\bibliographystyle{ACM-Reference-Format}
\bibliography{references}

%

\end{document}